\documentclass{aa}  
\usepackage{keyval}

\usepackage{natbib}
\bibpunct{(}{)}{;}{a}{}{,} 

\usepackage{graphicx}
\usepackage{caption}
\usepackage{txfonts}
\usepackage[]{hyperref}
\hypersetup{colorlinks,linkcolor={blue},citecolor={blue},urlcolor={blue}}

\begin{document} 

   \title{Fifteen new millisecond pulsars in 47~Tucanae}
   \titlerunning{Fifteen new millisecond pulsars in 47~Tucanae}
   \authorrunning{W.~Chen, D.~Risbud}
   \author{W.~Chen\thanks{\email{wchen@mpifr-bonn.mpg.de}}
          \inst{1},
          D.~Risbud
          \inst{1, 2, 7},
          P.~C.~C.~Freire
          \inst{1},
          A.~Ridolfi
          \inst{2},
          E.~Barr
          \inst{1},
          M.~Kramer
          \inst{1, 3},
          B.~Stappers
          \inst{3},
          \\
          F.~Camilo
          \inst{5},
          F.~Abbate
          \inst{1, 4},
          A.~Possenti
          \inst{4},
          Y.~P.~Men
          \inst{1},
          P.~V.~Padmanabh
          \inst{8,9},
          S.~M.~Ransom
          \inst{6},
          L.~Vleeschower
          \inst{3},
          \\
          V.~Venkatraman~Krishnan
          \inst{1},
          D.~J.~Champion
          \inst{1},
          Rene Breton
          \inst{3},
          V.~Balakrishnan
          \inst{10,1},
          S.~Buchner
          \inst{5}
          }

   \institute{Max-Planck-Institut f\"{u}r Radioastronomie, Auf dem H\"{u}gel 69, D-53121 Bonn, Germany
   \and
   Fakult\"at f\"ur Physik, Universit\"at Bielefeld, Postfach 100131, D-33501 Bielefeld, Germany
   \and
   Jodrell Bank Centre for Astrophysics, Department of Physics and Astronomy, The University of Manchester, Manchester M13 9PL, UK
   \and
   INAF -- Osservatorio Astronomico di Cagliari, Via della Scienza 5, I-09047 Selargius (CA), Italy
   \and
   South African Radio Astronomy Observatory, Cape Town, South Africa
   \and
   National Radio Astronomy Observatory (NRAO), 520 Edgemont Rd., Charlottesville, VA 22903 USA
   \and
   Rheinische Friedrich-Wilhelms-Universität Bonn, Regina-Pacis-Weg 3, D-53113 Bonn, Germany
   \and
Max Planck Institute for Gravitational Physics (Albert Einstein Institute), D-30167 Hannover, Germany
\and
Leibniz Universit\"{a}t Hannover, D-30167 Hannover, Germany 
\and
Center for Astrophysics | Harvard \& Smithsonian, Cambridge, MA 02138-1516, USA
   }
              
   \date{Received September xx, xxxx; accepted March xx, xxx}

 
  \abstract
   {47~Tucanae is one of the largest, brightest, and closest globular clusters to Earth. It hosts an exotic stellar population with stellar dynamics that indicate a complex evolution history. The cluster contains a large number of X-ray binaries and millisecond pulsars.} 
   {Searching for pulsars in this cluster has been an ongoing effort over more than two decades with fruitful results. However, given its large distance relative to the known pulsar population, previous surveys have found only the very brightest sources. Therefore, surveys with increased sensitivity should find many additional pulsars. Increasing the number of pulsars is crucial to investigate the dynamics of this globular cluster and could also lead to the discovery of unusual types of system.}
   {With a significantly increased sensitivity compared to earlier telescopes, MeerKAT is the natural choice to perform new surveys. We carried out two campaigns with different observational cadences to account for the high scintillation along the line of sight to this cluster. All observations were carried out using multi-beam filter-banking user-supplied equipment to increase efficiency and localisation.}
   {Here we report the discovery of fifteen new pulsars in 47~Tucanae with MeerKAT. These discoveries bring the total number of known pulsars in this globular cluster to 42, and the MeerKAT discoveries in this cluster to 17. We discuss some of their characteristics, which include preliminary localisations and estimates of orbits for most systems. Highlights include the discovery of 47~Tuc~af, a 'black widow' pulsar with a short orbital period that was identified optically in 2002 as a candidate binary pulsar, and 47~Tuc~ai, an eccentric binary pulsar with a massive companion, a unique system in 47~Tuc to date. Apart from the new systems, we also re-detect and localise 47~Tuc~P and V, two elusive, seldom-detected systems that had no precise localisation from a phase-connected timing solution. The localisation of 47~Tuc~V places it in a position consistent with a continuum source detected earlier in MeerKAT imaging data.}
   {}


   \maketitle
%

\section{Introduction}\label{sec:intro}

   Exotic pulsar systems have been used to test fundamental physics, such as testing general relativity \citep{Voisin2020,Kramer2021,FreireWex2024} and constraining the equation of state of dense matter, which depends on the characteristics of the strong nuclear force \citep{OzelFreire2016}, especially through the measurement of large neutron star (NS) masses \citep{Fonseca2021}. They are also extremely valuable for improving our understanding of stellar evolution in binary systems \citep{Tauris2023} and many other astrophysical applications.

   Most pulsars are found to be associated with the Galactic disc. However, the high stellar densities (which frequently exceed one million solar mass per cubic pc; see \citealt{Baumgardt2018}) in globular clusters (GCs) lead to close stellar encounter rates that are many orders of magnitude larger than those found in most of the Galaxy \citep[$\approx$~0.1~pc$^{-3}$,][]{Chabrier2003}. This leads to exchange encounters where many otherwise undetectable neutron stars are paired with main sequence stars. The latter's evolution eventually leads to the formation of low-mass X-ray binaries at a rate three orders of magnitude larger than that in the Galactic disc \citep{Clark1975}. Many of these systems eventually evolve into radio millisecond pulsars (MSPs), which are also several orders of magnitude more numerous per unit of stellar mass than those in the Galactic disc: thus far, 362 pulsars have been discovered in 46 globular clusters\footnote{\url{https://www3.mpifr-bonn.mpg.de/staff/pfreire/GCpsr.html}}, most of them are MSPs, compared with a total of 552 MSPs in the entire galactic disc\footnote{\url{https://pages.astro.umd.edu/~eferrara/GalacticMSPs.html}}.

   Not all GCs form low-mass X-ray binaries (LMXBs) and MSPs equally: some GCs (especially the denser, larger ones) have a higher rate of stellar interactions \citep[$\Gamma$,][]{Verbunt1987,Bahramian2013}. As expected, GCs with large $\Gamma$ tend to have larger X-ray binary and MSP populations.
   
   However, the size of the pulsar population in a GC is not the only important factor. If the total interaction rate is high relative to the number of stars \citep[i.e. a large value of $\gamma \equiv \Gamma / n_{*}$,][]{Verbunt2014}, then even after being recycled in an LMXB (which in GCs already formed in exchange interactions), an individual NS still has a sizeable probability of undergoing subsequent exchange encounters, either during the recycling phase or after its completion. These 'secondary exchange encounters' lead to the formation of extremely unusual systems, such as possible MSP-black hole binaries like PSR~J0514$-$4002E in the GC NGC~1851 \citep{Barr2024}. Such systems can be particularly useful for the fundamental physics studies mentioned above; their discovery remains a strong motivation for pulsar surveys in GCs. These systems occur, as expected, only in GCs with large values of $\gamma$.
   
   The globular cluster NGC 104, or alternatively 47~Tucanae (47~Tuc from here on), is located in the Southern Hemisphere at equatorial sky coordinates  $\alpha = 00^\mathrm{h}\,24^\mathrm{m}\,05\fs67$, $\delta = -72^\circ\,04'\,52\farcs6$, or Galactic coordinates $l = 305.89^\circ$, $b = -44.89^\circ$ at 4.69 kpc away from Earth \citep{2012AJ....143...50Woodley}.
   With a tidal radius of 43\farcm8 \citep{Shao2019}, it is one of the largest GCs associated with our Galaxy and is the second brightest after Omega Centauri. As a consequence, it is one of the best studied globular clusters. This is also the case for extensive radio surveys, which prior to this work discovered a total of 27 pulsars.

   Due to the southern position of this cluster, all such surveys until 2017 were carried out with the Murriyang 64-m radio telescope near Parkes, NSW \citep{Manchester1990,Manchester1991,Robinson1995,Camilo2000,Knight2007,Pan2016,Freire2017}. These surveys resulted in a total of 25 pulsar discoveries, ten of which are isolated pulsars, and the remaining 15 are binaries. They have a narrow range of spin periods (2.1 to 7.6 ms) and a narrow range of dispersion measures (DMs), from 24.2 to 24.9 pc cm$^{-3}$.

   The discovery and follow-up of these pulsars was scientifically very fruitful.
   Timing studies enabled studies of the mass model of the cluster, mass segregation \citep{Freire2001,Freire2003,Pan2016,Freire2017}, and the detection of ionised gas in the cluster \citep{Freire2001b,Abbate2018}.
   The precise timing localisations allowed for an extensive study of these pulsars and their systems across the electromagnetic spectrum. All pulsars were detected in X-rays \citep{Grindlay2001,Bogdanov2005, Bogdanov2006,Bhattacharya2017,Hebbar2021}; this also allowed measurements of the intra-shock regions of some of the binaries, especially 47~Tucanae~W (henceforth '47~Tuc~W', and likewise for all other pulsars). At optical wavelengths, this led to the detection of the redback companion of 47~Tuc~W \citep{Edmonds2002} in archival \textit{Hubble} Space Telescope (HST) data and the detection of several He WD companions of several MSPs \citep{Edmonds2001,Rivera-Sandoval2015,Cadelano2015}.
   
   The MeerKAT telescope has resulted in an order of magnitude increase in sensitivity for radio sources in the southern sky compared to the Murriyang telescope \citep{Jonas2016}; this sensitivity implies that searches for pulsars in GCs with MeerKAT have great potential for many new discoveries. Thus far these searches have been mostly carried out in a collaboration between TRAnsients and PUlsars with MeerKAT \citep[TRAPUM, ][]{StappersKramer2016} and MeerTime
   \citep{Bailes2020} Large Science Projects. MeerTime has provided observing time as part of its 'Relativistic Binaries' topic, as there were previously known relativistic binaries in GCs for which dedicated timing was fruitful \citep[see][]{Kramer2021b}.
   With a current total of 121 discoveries in GCs\footnote{\url{https://trapum.org/discoveries/}} \citep{Ridolfi2021,Douglas2022,Ridolfi2022,Vleeschower2022,Abbate2022,Zhang2017,Chen2023,Abbate2023b,Vleeschower2024,Padmanabh2024}, this has been the most successful search for pulsars in GCs to date. 
   
   In the early search for pulsars in GCs with MeerKAT, \cite{Ridolfi2021} discovered two new pulsars in 47~Tuc, namely 47~Tuc~ac and 47~Tuc~ad, in data taken using the L-band receivers with the Pulsar Timing User-Supplied Equipment \citep[PTUSE;][]{Bailes2020}. In this work, we present further pulsar discoveries in 47~Tuc made as part of a targeted survey, especially in the UHF band (544-1088 MHz).
   
   This paper is organised as follows. We describe our observations of 47~Tuc including the data reduction details in Sects.~\ref{sec:observations}~and~\ref{sec:parkes_redetections}, new discoveries and their fundamentals in Sect.~\ref{sec:discoveries}. Then, in Sects.~\ref{sec:discussion}~and~\ref{sec:summary}, we discuss the implications of these discoveries.

\section{Observations}
\label{sec:observations} 

\subsection{Two different observing campaigns}
\label{chap:observational_campaign}
To tackle the effect of scintillation screens, different observational strategies were used in two campaigns; the dates and durations of the observations are shown in Tables \ref{tab:observation_list_1} and \ref{tab:observation_list_2}. Both campaigns cast numerous coherently synthesized beams (CBs) to cover the cluster. 

Ten observations were carried out in the first campaign, which spanned from May 2020 to February 2021. The longest interval between observations was 104 days, and the shortest was seven days. The first observation, 10L, lasted four hours; the remaining observations lasted approximately one hour.  
In this campaign, five observations were conducted in the L band (856–1712 MHz) and the other five were recorded in the UHF band (544–1088 MHz). Similar antenna configurations, with 56 antennas from both the core and the outer array, were used throughout these observations. However, different epochs of the observations resulted in variations in the synthesized beam shapes\footnote{It should be noted that the version of \textsc{Mosaic} running on FBFUSE is different between these two campaigns. Particularly, the newer version \citep[described by][]{chen2021} uses a significantly different approach to characterize the shape of the beams. These changes result in different tiling sizes for the same overlap ratio between CBs between the two campaigns.} \citep{chen2021}, therefore, the total sky coverage of the tilings of CBs varies between observations. In observation 11L, for example, the distance of the furthest CB to the cluster centre is 6.75 arcminutes, while it is 14.82 arcminutes for observation 12L, corresponding to nearly five times increase in covered area.

The second campaign included an exceptionally long observation of more than 19 hours in total. This long observation was split into multiple closely adjoined time segments of different lengths, such as approximately one, two, three, four, and five hours. More details about this campaign can be found in \cite{Abbate2023a}.
All of these long-integration observations were recorded in the UHF band. Unlike in the first campaign, the number of CBs is twice as small because a higher time resolution was used. Thus, with a similar beam size, the sky coverage of the CB tilings were much smaller (roughly 1 arcminute, slightly smaller than the zoomed-in area shown in Fig.~\ref{fig:15L_tiling}). 

The raw responses of the individual dishes were converted to digital raw voltages on the dish and transferred to the Karoo Array Processor Building. The geometric integer delays and instrument coarse delays were applied to these raw voltages before they were distributed to the F-engines that perform channelisation. Then, geometric fine delays were applied to the channelised voltages to synchronise the phases across the array. Afterwards, they were broadcast to the MeerKAT internal network. 

Filter-Banking User-Supplied Equipment \citep[FBFUSE;][]{Barr2018} receives these phase-aligned data, applies complex instrument delays obtained from the MeerKAT Science Data Processor (SDP), then generates extra geometric weights using \textsc{Mosaic}\footnote{\url{https://github.com/wchenastro/Mosaic}} and applies them to the data across all channels. After that, FBFUSE forms a collection of off-boresight CBs by summing up the corresponding weighted data. 

After an observation, depending on the requirements of the scientific application, FBFUSE further sub-bands the beamformed data to reduce the number of channels to alleviate the stress of the storage pool. In this process, the data were incoherently de-dispersed according to the value of a DM, the integrated column density of free electrons in the interstellar medium between the source and the observer. The value of the DM used in the process is 24.39 pc cm$^{-3}$, which is the average DM of known pulsars of 47~Tuc at the time of observation, and then the adjacent channels were summed. For each CB, the time resolution was 76.561 $\mu$s in the L-band and 60 $\mu$s in the UHF band. Later, these data were brought to Germany on hard drives and loaded into the Hercules computing cluster\footnote{\url{https://docs.mpcdf.mpg.de/doc/computing/clusters/systems/Radioastronomy.html}}.

\subsection{Data reduction}

The pulsar search pipeline starts with Radio Frequency Interference (RFI) cleaning of data using \texttt{filtool} from \cite{Men2023}. This tool calculates the kurtosis and skewness of the data to identify RFI and replace it with the mean value of the data. Then, a search routine was supervised by \textsc{Pulsar\_Miner}\footnote{\url{https://github.com/alex88ridolfi/PULSAR_MINER}} \citep{Ridolfi2021}, utilising various programs from the pulsar search and analysis toolkit \textsc{PRESTO}\footnote{\url{https://github.com/scottransom/presto}} \citep{Ransom2011}.
It used \texttt{rfifind} to flag RFIs and created a DM trial scheme using \texttt{DDplan.py} in the range between 23.5 and 25.5 with a step of 0.05~pc~cm$^{-3}$. Then it searched for periodic signals using \texttt{accelsearch} in data chunks of different lengths (i.e. 10\,m, 20\,m, 30\,m, 60\,m, and full length) to cover pulsars with a wide range of periods \citep{Ransom2003}, and fold the candidates with \texttt{prepfold}. The resulting candidates went through a sifting process based on their signal-to-noise ratio (S/N) and were cross-checked against the harmonics of known pulsars. An orchestration pipeline that manages the job submission and cascade process was created to fully automate this search routine in the cluster. The high-ranking candidates were then bundled for human inspection. A more detailed description of this pipeline can be found in \cite{Chen2023}.

\subsection{Sensitivity}
The minimum detectable flux density (S$_{\text{min}}$) of these two campaigns was calculated using the modified radiometer equation given by \cite{Dewey1985}. The values of the parameters used in the equation and S$_{\text{min}}$ in the L band (effective bandwidth: 640 MHz) are similar to those reported for other TRAPUM observations \citep[e.g.][]{Chen2023}.  Because the observations in this work have variable lengths, the sensitivity is scaled by them. Typically, for a detection with an S/N of ten, a four-hour observation has S$_{\text{min}}$ of 12.8 $\mu$Jy and a one-hour observation has S$_{\text{min}}$ of 25.5 $\mu$Jy, assuming 56 antennas in the UHF band (effective bandwidth: 435 MHz). 

However, the scintillation significantly changes the flux densities of radio pulsars in this globular cluster on timescales of $\sim$\,15 minutes to one hour. Thus, some of the fainter pulsars will only be detected rarely, at times when interstellar scintillation amplifies their flux density as seen at the telescope by a large amount. This constantly varying flux density makes it hard to estimate the 'average' flux densities for newly discovered pulsars, especially those that are detected infrequently.

\begin{table}[h]
    \centering
    \caption{Observation list of the first campaign.}
    \begin{tabular}{cccc}
        \hline \hline
        Index & Date & Duration (h) & Band \\
        \hline
        10L & 2020-05-02 & 4.0 &  L \\
        11L & 2020-07-29 & 0.8 &   L   \\
        12U & 2020-11-10 & 0.8 &  U  \\
        14L & 2020-11-21 & 1.0 &  L    \\
        15L & 2020-12-17 & 0.8 &  L   \\
        17L & 2021-01-14 & 0.8 &  L   \\
        18U & 2021-01-22 & 1.0 &  U  \\
        19U & 2021-01-30 & 1.0 &  U  \\
        20U & 2021-02-06 & 1.0 &  U \\
        21U & 2021-02-17 & 1.0 &  U  \\
        \hline
    \end{tabular}

\tablefoot{The observations carried out with L-band receivers are denoted with a letter 'L'; similarly, those carried out with UHF-band receivers are denoted with a letter 'U'.}
    \label{tab:observation_list_1}
\end{table}

\begin{table}[h]
    \centering
    \caption{Observation list of the second campaign.}
    \begin{tabular}{lcr}
        \hline \hline
        Index & Date \& Start time & Duration (h) \\
        \hline
        25U  & 2022-01-26 15:32:32 & 2.0  \\
        26U1 & 2022-01-27 05:10:30 & 1.9  \\
        26U2 & 2022-01-27 07:22:40 & 5.0 \\
        26U3 & 2022-01-27 14:09:21 & 0.9  \\
        26U4 & 2022-01-27 15:18:31 & 3.0 \\
        26U5 & 2022-01-27 18:32:35 & 3.5 \\
        26U6 & 2022-01-27 22:10:57 & 1.0  \\
        27U  & 2022-01-28 05:10:37 & 1.9  \\
        28U  & 2022-01-29 07:08:31 & 2.0  \\
        \hline
    \end{tabular}
\tablefoot{All observations in this campaign were carried out with UHF receivers.}
    \label{tab:observation_list_2}
\end{table}

\subsection{Localisation}

For each new pulsar, or previously known pulsar without a known position, we folded the beams surrounding the one with the best S/N of the detection using an ephemeris derived from timing analysis from multiple observations, when available. For pulsars without an ephemeris, the folding parameters that produced the highest S/N in the best beam were used to fold the surrounding beams. We collected the S/N value from each folded beam and used it as input for \textsc{SeeKAT}\footnote{\url{https://github.com/BezuidenhoutMC/SeeKAT}} \citep{Bezuidenhout2023}, together with the point spread function (PSF) generated using \textsc{Mosaic}, to obtain the best-fit positions of the pulsars. The results of the localisation for the new pulsars are listed in Table \ref{tab:discovery_list}.
This technique allowed us to determine the positions of new sources without the need for additional observations, which is crucial for sporadic sources such as transitional pulsars and fast radio bursts, but also for strongly scintillating pulsars such as those in 47~Tuc. These fine positions enabled rapid, targeted follow-up observations of the new sources.

The localisation results for known pulsars that were previously unlocalised are listed in Table \ref{tab:localized_position_known_pulsar}. Note that the positions calculated by \textsc{SeeKAT} are highly influenced by the quality of the detections in the surrounding beams, which itself has been affected by the RFI situations of their directions. Furthermore, differences between the theoretical and actual PSF of the beams also affect the results. The positions will only be significantly improved (by many orders of magnitude) by timing these pulsars when more detections become available. During long observations, the on-sky beam shapes change as a function of time and the altitude of the source at the observing site, both of which are additional sources of uncertainty.

\begin{figure*}
    \centering
    \includegraphics[width=\textwidth]{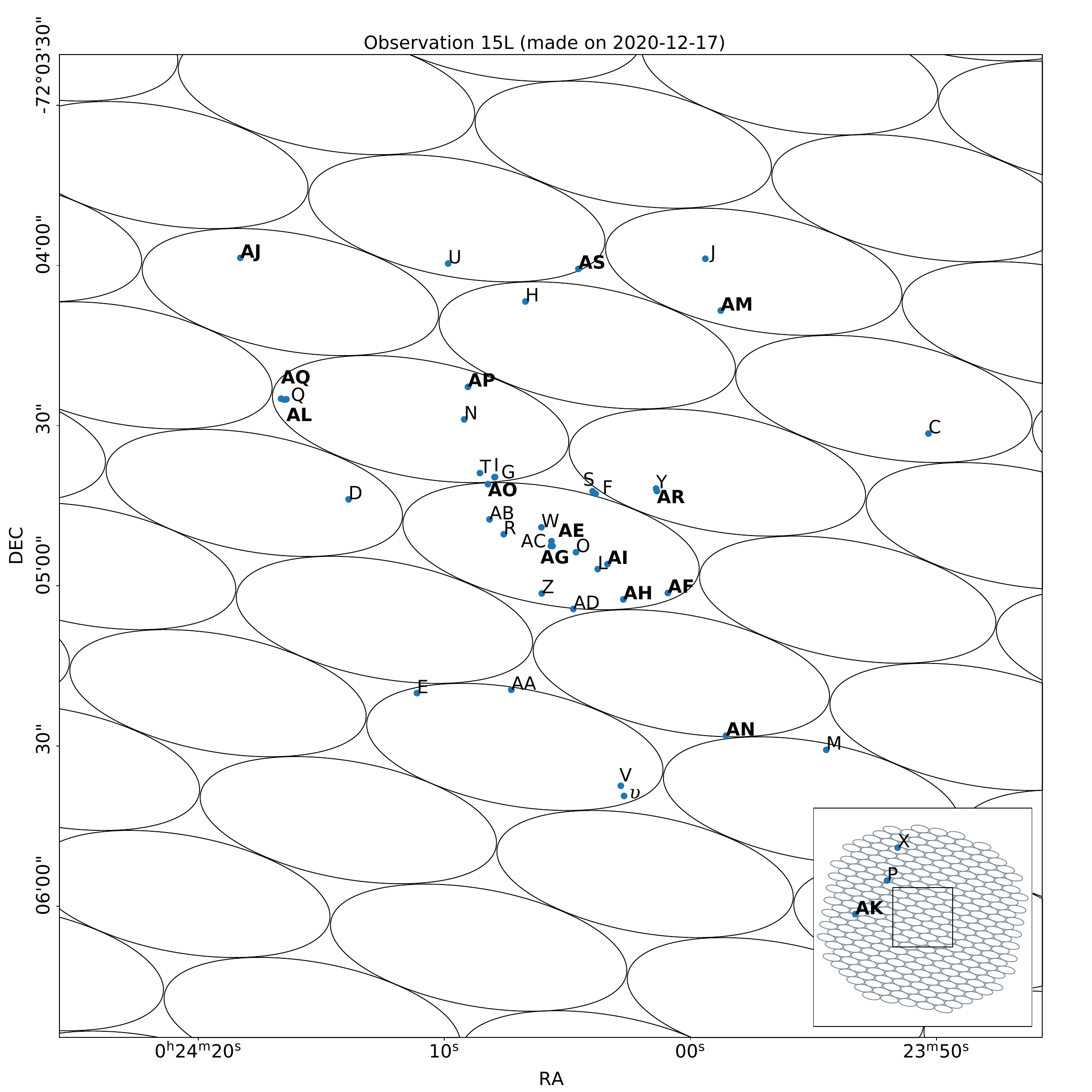}
    \caption{Positions of the pulsars overlaid on the positions of the MeerKAT beams from the observation on 2020-12-17 with the L-band receiver. The ellipses represent coherent synthesised beams with their edges indicating a 70\% gain level. The gaps between beams were still covered, albeit with lower gain. In the inset, which shows a broader view, the positions of 47~Tuc~P, X, and ak can be seen; these three pulsars lie more than 3 arcminutes from the centre of the cluster. The symbols of new discoveries from this work are in bold. The \textsc{$\upsilon$} symbol indicates the position of the source mentioned in \cite{Heywood2023}. }
    \label{fig:15L_tiling}
\end{figure*}

\begin{figure*}
    \centering
    \setlength\tabcolsep{1.0pt}
    \renewcommand{\arraystretch}{0.1}
    \begin{tabular}{llllllll}
    \includegraphics[width=0.14\textwidth]{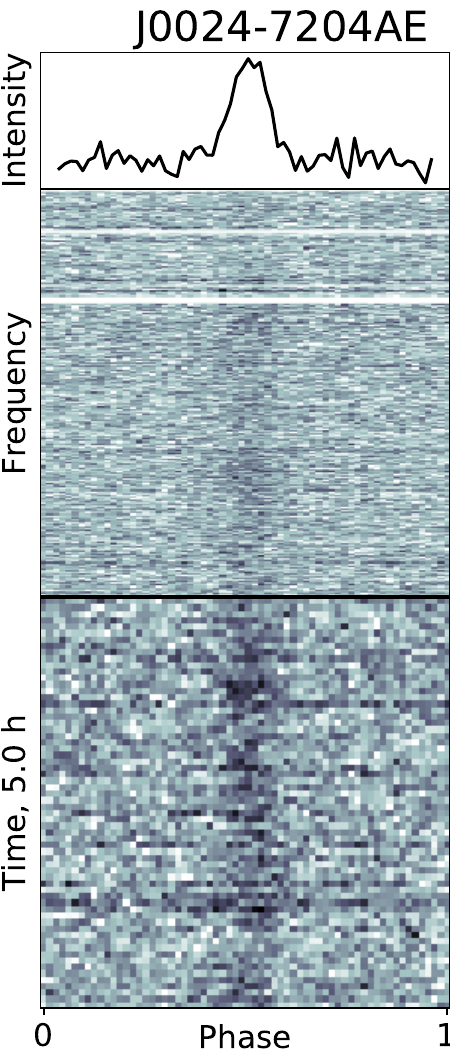} &
    \includegraphics[width=0.14\textwidth]{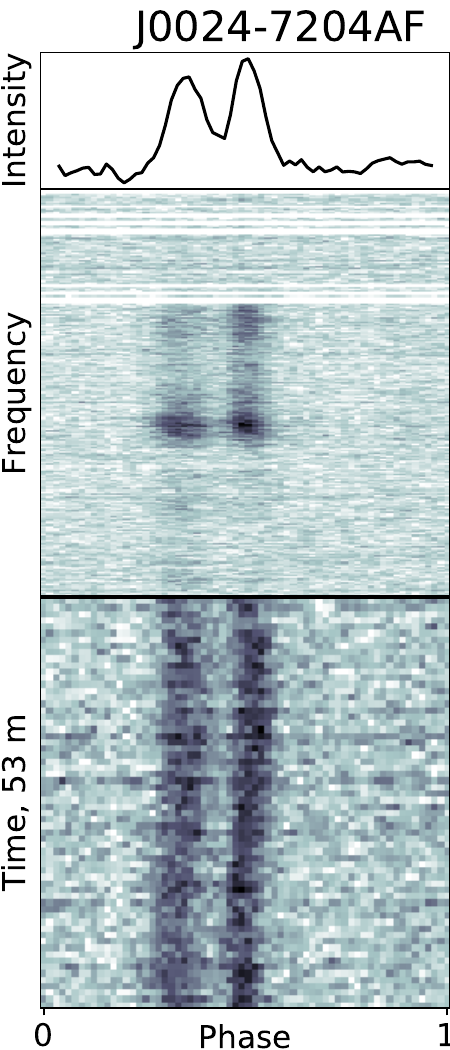} &
    \includegraphics[width=0.14\textwidth]{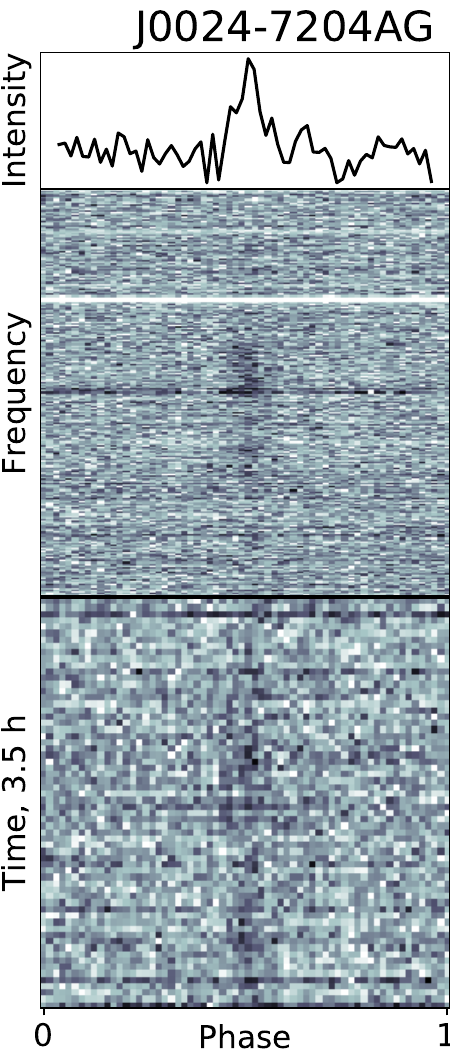} &
    \includegraphics[width=0.14\textwidth]{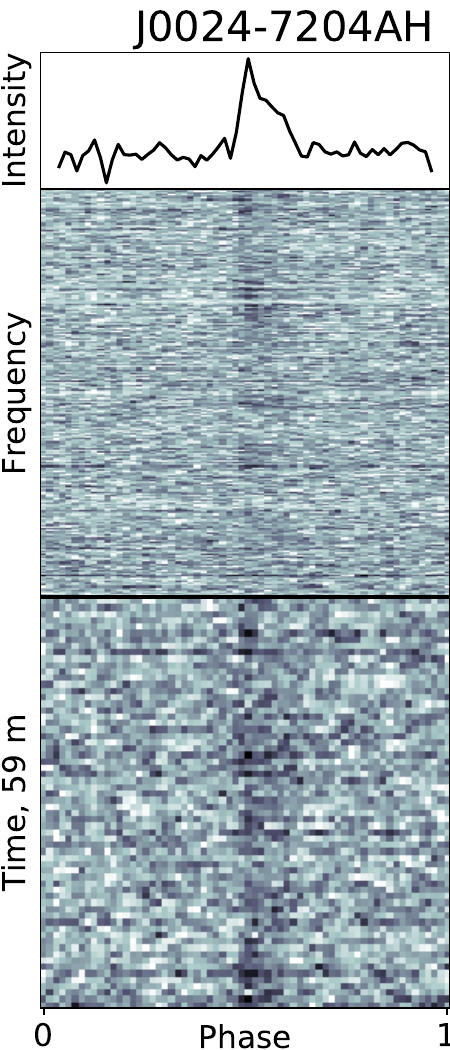} &
    \includegraphics[width=0.14\textwidth]{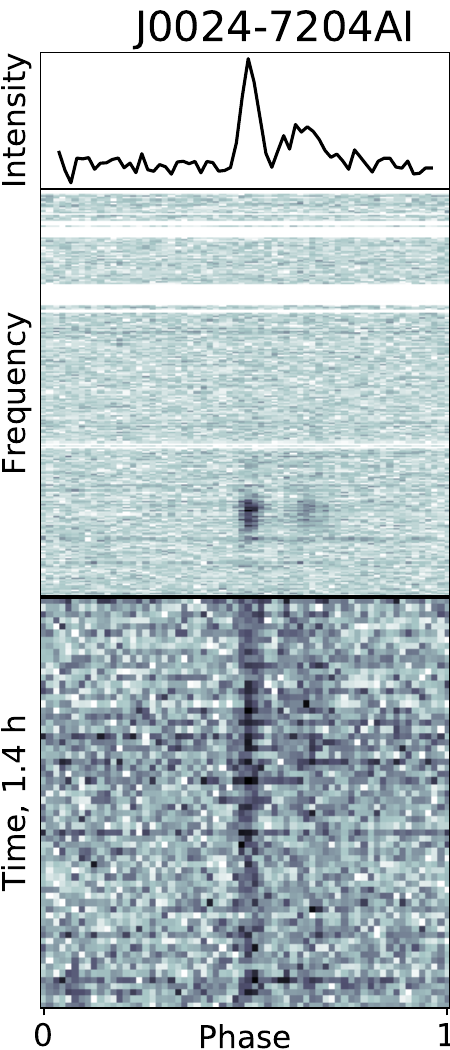} &
    \includegraphics[width=0.14\textwidth]{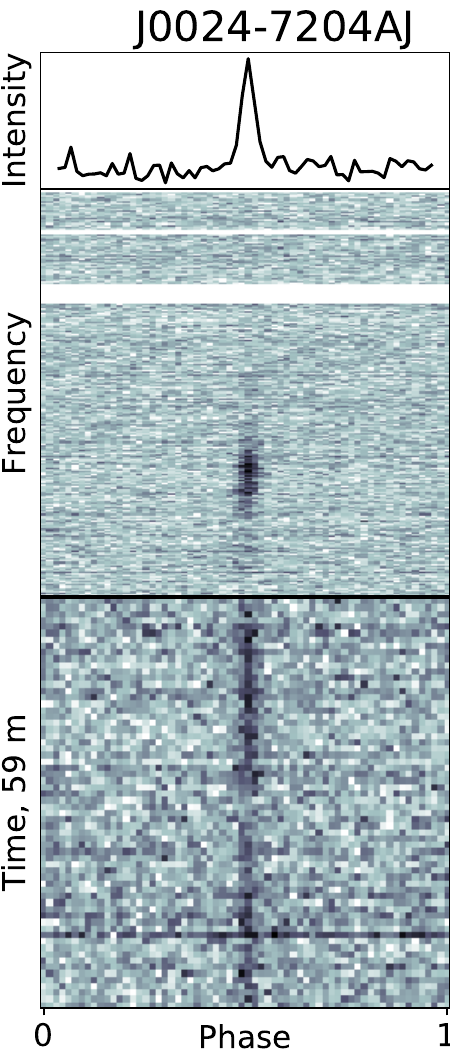} \\
    \includegraphics[width=0.14\textwidth]{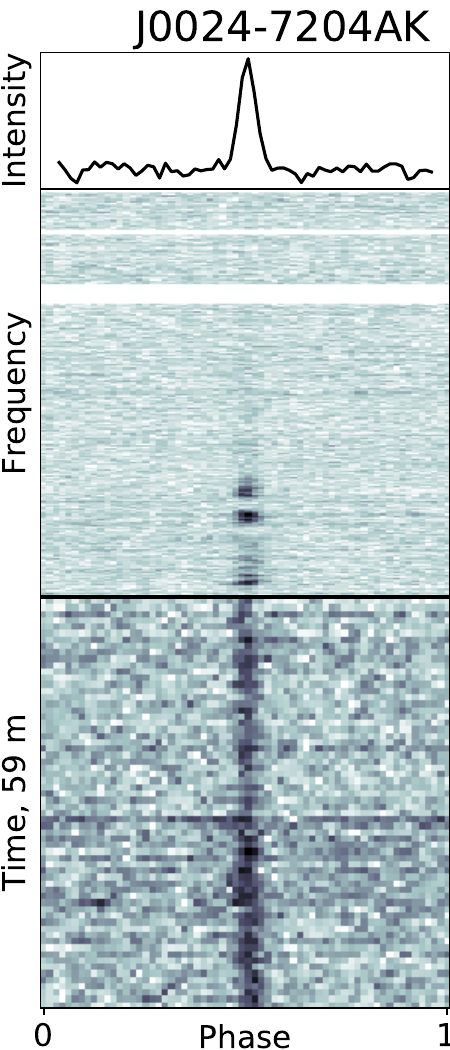} &
    \includegraphics[width=0.14\textwidth]{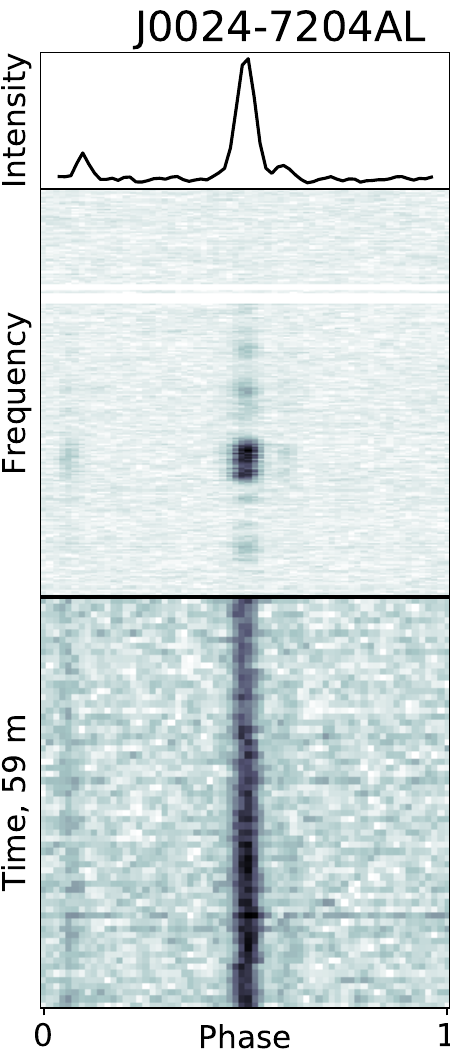} &
    \includegraphics[width=0.14\textwidth]{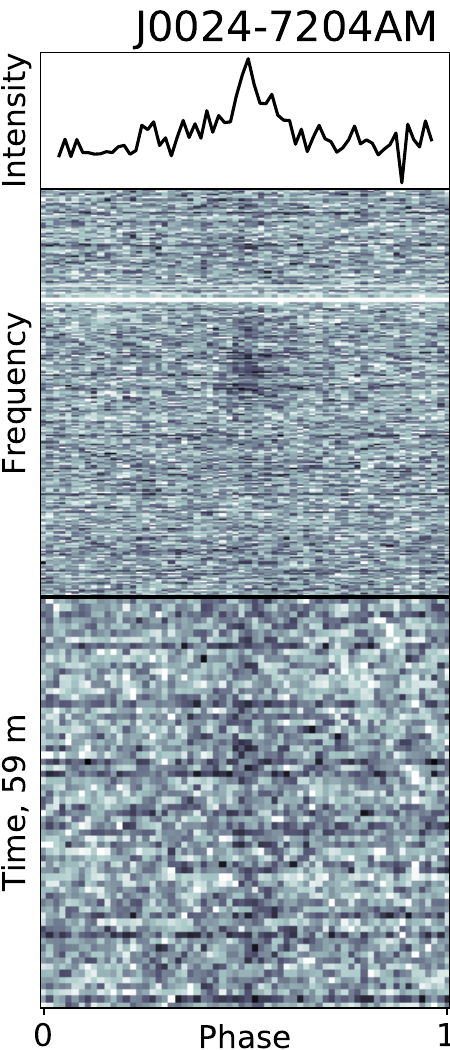} &
    \includegraphics[width=0.14\textwidth]{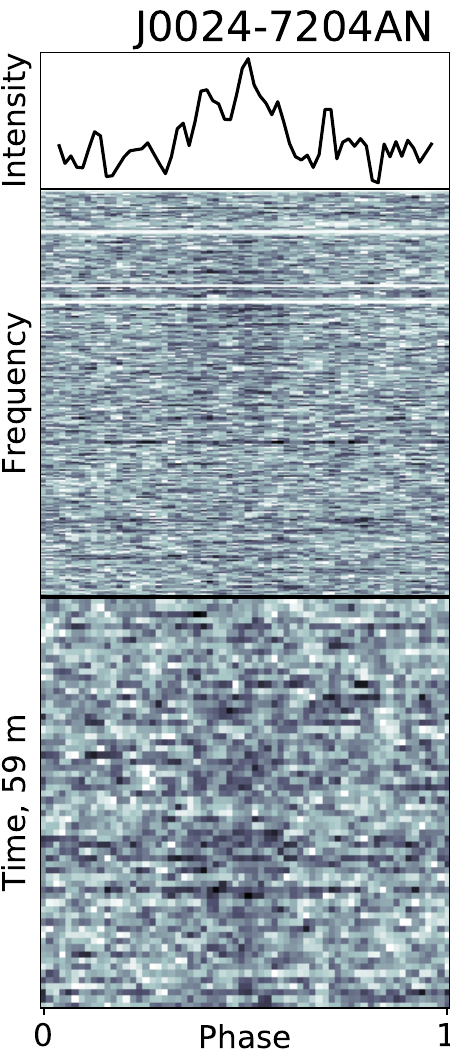} &
    \includegraphics[width=0.14\textwidth]{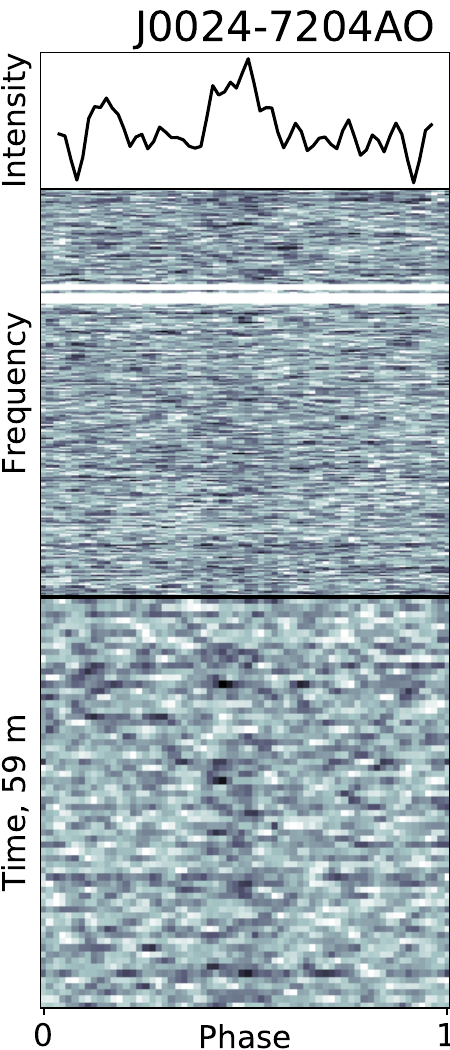} &
    \includegraphics[width=0.14\textwidth]{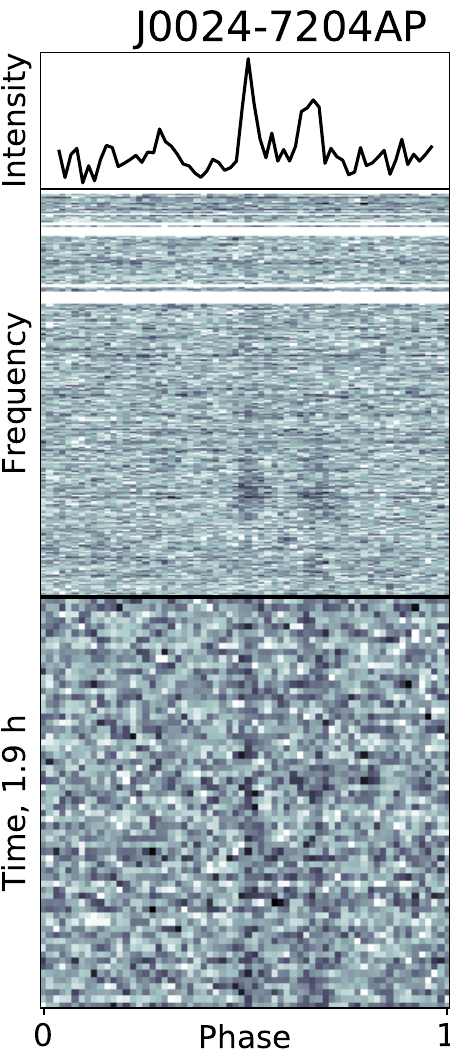} \\
    \includegraphics[width=0.14\textwidth]{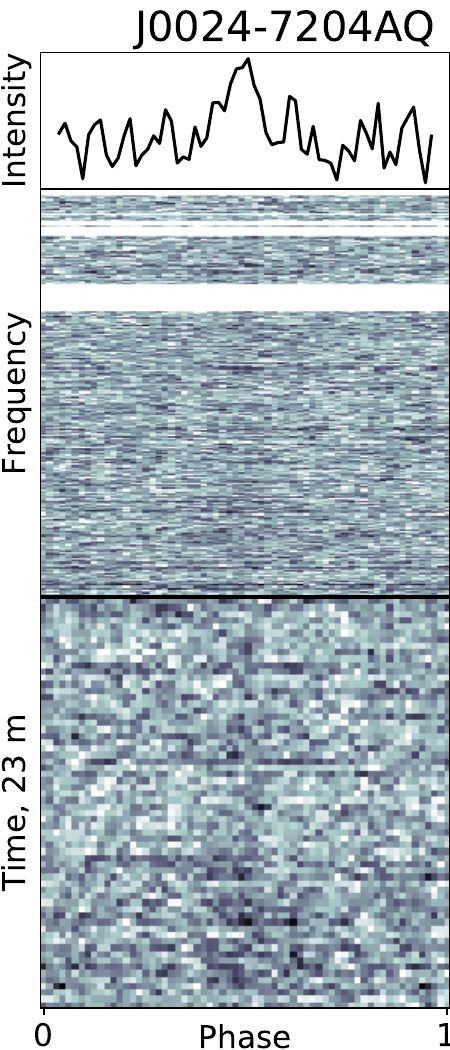} &
    \includegraphics[width=0.14\textwidth]{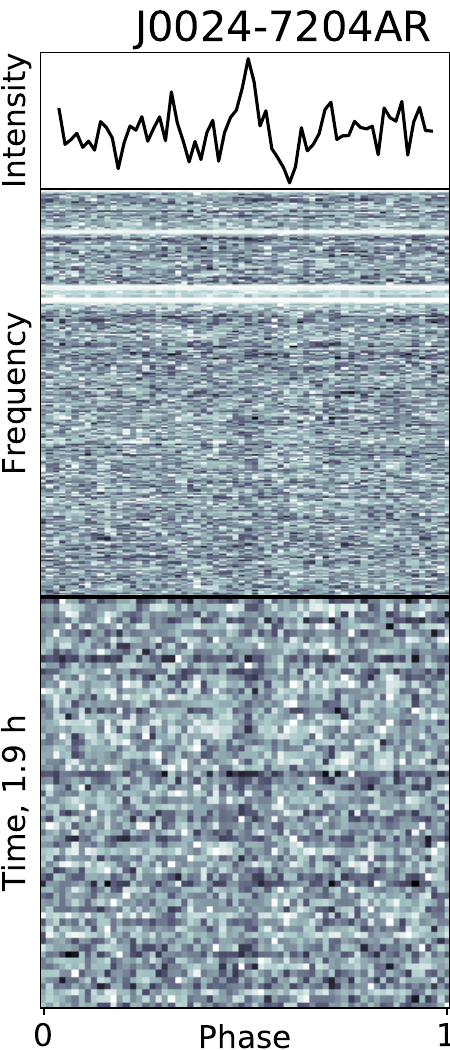} &
    \includegraphics[width=0.14\textwidth]{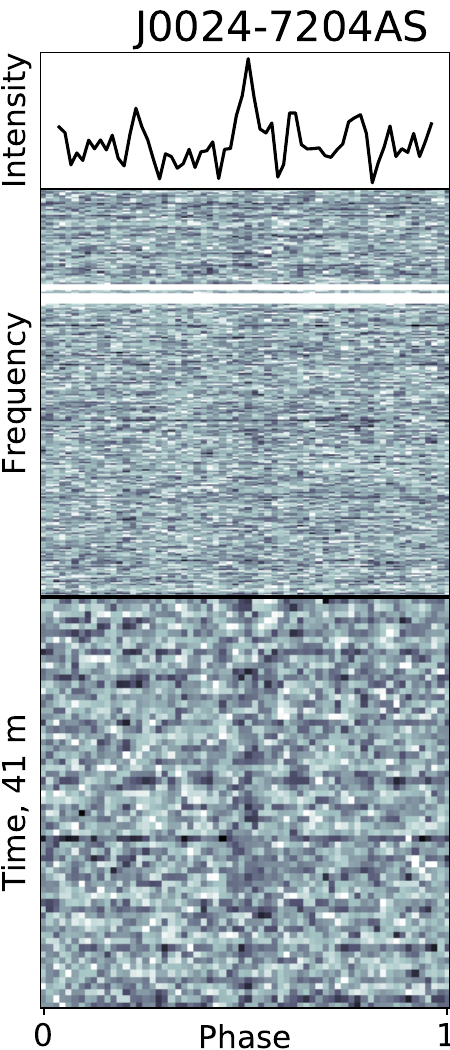} &
    
    \end{tabular}
    \caption{Profiles of the new pulsars from their highest S/N detections. The Y-axis is intensity, frequency, and time (with duration) from top to bottom panels. The X-axis is the phase window from 0 to 1 with 64 phase bins and all three panels of each pulsar share the same X-axis. All of these profiles are folded with data from UHF observations with the same sub-banding of 256, thus every frequency-versus-phase panel has a frequency resolution of 2.125 MHz.}
    \label{fig:profiles_of_new_pulsars}
\end{figure*}

\section{Discoveries}
\label{sec:discoveries}
From these two campaigns, we discovered 15 new pulsars. According to preliminary assessments, three are isolated and the remaining 12 are in binary systems. Notably, 47~Tuc~ai has an eccentric orbit and 47~Tuc~af has a distinct profile change between the L and UHF bands. The fundamental parameters, positions, and pulse profiles of these pulsars are listed in Table \ref{tab:discovery_list} and Figs.~\ref{fig:15L_tiling} and \ref{fig:profiles_of_new_pulsars}.

\begin{table*}[h]
    \centering
    \caption{Discoveries in this work. }
    \begin{tabular}{llrllllll}
         \hline \hline
          PSR & Type     &  $P$    &  DM            & $P_b$ & $x_p$  & $M^{\textrm{min}}_c$ & $\alpha$ & $\delta$  \\
                  &          &   (ms)  & (pc cm$^{-3}$) &  (d)  & (lt-s) & M$_\odot$  & J2000 & J2000 \\  \hline
          ae & He WD & 3.87 & 24.34 & 0.757 &  0.741 & 0.121 & 00:24:05(1) & $-$72:04:51(2) \\
          af & BW/RB & 2.99 & 24.34 & 0.0677 &  0.0852 & 0.0676 & 00:24:04(1) & $-$72:04:51(1) \\
          ag & Binary & 9.76 & 24.41 & 1.08 & 0.0821 & 0.0101 & 00:24:05(1) & $-$72:04:52(21)$^{\ddag}$  \\
          ah & Binary & 3.07 & 24.36 & - & - & - & 00:24:02(1) & $-$72:05:02(17)$^{\ddag}$ \\
          ai & Massive & 13.03 & 24.47 & 1.65 & 5.35 & 0.628 & 00:24:03(4) & $-$72:04:55(7) \\
          aj & Isolated & 6.36 & 24.38 & i &  i & i & 00:24:18(2) & $-$72:03:57(9) \\
          ak & Binary & 3.52 & 23.91 & - & - & - & 00:24:50(2) & $-$72:04:42(6) \\
          al & BW & 2.67 & 24.11 & 0.157 &  0.0206 & 0.00910 & 00:24:41(5) & $-$72:05:00(77)$^{\ddag}$ \\
          am & Binary & 4.16 & 24.55 & - &  - & - & 00:23:58(6) & $-$72:04:08(21) \\
          an & Binary & 2.61 & 24.12 & - &  - & - & 00:23:58(2) & $-$72:05:28(14) \\
          ao & Isolated & 1.88 & 23.65 & i &  i & i & 00:24:08(4) & $-$72:04:41(34) \\
          ap & Isolated & 5.11 & 24.36 & i &  i & i & 00:24:09(2) & $-$72:04:22(9) \\
          aq & Binary & 3.04 & 23.63 & - &  - & - & 00:24:16(18) & $-$72:04:24(30) \\
          ar & Binary & 9.76 & 24.16 & - &  - & - & 00:24:01(15) & $-$72:04:42(85) \\
          as & Binary & 4.02 & 24.66 & - &  - & - & 00:24:04(7) & $-$72:04:00(26) \\
           \hline
    \end{tabular}

    \tablefoot{The 2$\sigma$ credible interval of the position are given by \textsc{SeeKAT}. The symbol $\ddag$ indicates that the pulsar does not have significant multi-beam detection to further localise its position. For this reason, the width (major axis) of the synthesised beam is used for the errors. Likewise, their positions are represented by the coordinates of the centre position of the beams in which they were detected with the highest S/N. The binary pulsars are listed as 'Binary' if the orbital parameters have not yet been determined, otherwise they are listed as millisecond pulsar - Helium white dwarf (He WD) binaries,
    as 'black widows' (BW), 'redbacks' (RB), or, in the case of Tuc ai, as 'massive' (see discussion). The minimum companion masses are calculated assuming a pulsar mass of $M_{\rm p} = 1.4 \, \rm M_{\odot}$.}
    \label{tab:discovery_list}
\end{table*}

\subsection{Isolated pulsars}
\subsubsection{47~Tuc~aj}
47~Tuc~aj was detected in both campaigns near the centre of the cluster, though only in the UHF band. It has a spin period of 6.36 ms and a DM of 24.37 pc cm$^{-3}$. Its period varies insignificantly between observations and all detections can be recovered with the same ephemeris. For this reason, we conclude, preliminarily, that it is an isolated pulsar.

\subsubsection{47~Tuc~ao}
47~Tuc~ao was detected in only one observation (26U6). This pulsar differs from the general pulsar population of 47~Tuc by having a relatively short spin period of 1.88 ms. Both 47~Tuc~ao and 47~Tuc~aq have significantly lower DMs of 23.64 pc cm$^{-3}$ compared to other pulsars in the cluster. It is considered a confirmed discovery due to its detection in multiple surrounding beams, which was also useful for its localisation. However, the precision of the position is low because of the low S/N. The spin period shows no detectable variation between detections; for this reason, we conclude, preliminarily, that it is an isolated pulsar or potentially a binary with a long orbit. The discovery of a pulsar with a DM at the lower edge of the search range prompts a relaxation of the criteria for future searches of this cluster. 

\subsubsection{47~Tuc~ap}
47~Tuc~ap was detected in the long-integration campaign near the centre of the cluster in the UHF band. The detection is from a dedicated beam pointed to the known pulsar 47~Tuc~N, however its spin period (5.11 ms) and DM (4.34 pc cm$^{-3}$) distinguish it from that pulsar. Detections from surrounding beams were subsequently obtained using the same folding parameters. Besides detections from 27U and 28U, no additional detection was found after folding the data from the rest of the long-integration campaign or the first campaign. No significant period variation was observed in two epochs with a separation of approximately one day, so we consider it as an isolated pulsar or a binary with a long orbit.

\subsection{Binary pulsars}

The remaining 12 discoveries are apparently in binary systems, judging from the variations in their observed spin periods between different observations or within a single observation. For discoveries with detections in multiple epochs, we extracted their spin periods, with which we tried to derive rudimentary orbital solutions using the method described by \cite{Freire2001}. Then we used the program \texttt{fit\_circular\_orbit.py} from \textsc{presto} with those solutions to obtain initial ephemerides. With these ephemerides, we performed a timing analysis of the data to improve their orbital solutions with \textsc{tempo}\footnote{\url{https://sourceforge.net/projects/tempo/}} or \textsc{tempo2}\footnote{\url{https://bitbucket.org/psrsoft/tempo2/src/master/}} \citep{Hobbs2006}. In some cases, the ephemerides were used with \textsc{spider\_twister}\footnote{\url{https://alex88ridolfi.altervista.org/pagine/pulsar_software_SPIDER_TWISTER.html}} to search the observations for additional detections, which were then used to further improve the ephemerides. Since a large portion of the newly discovered binary pulsars in this work have only sporadic or weak detections, their current ephemerides are not tightly constrained and will require improvement when more observations become available.

\subsubsection{47~Tuc~ae}\label{sec:47Tuc_ae}
\begin{figure}
    \centering
    \includegraphics[width=0.48\textwidth]{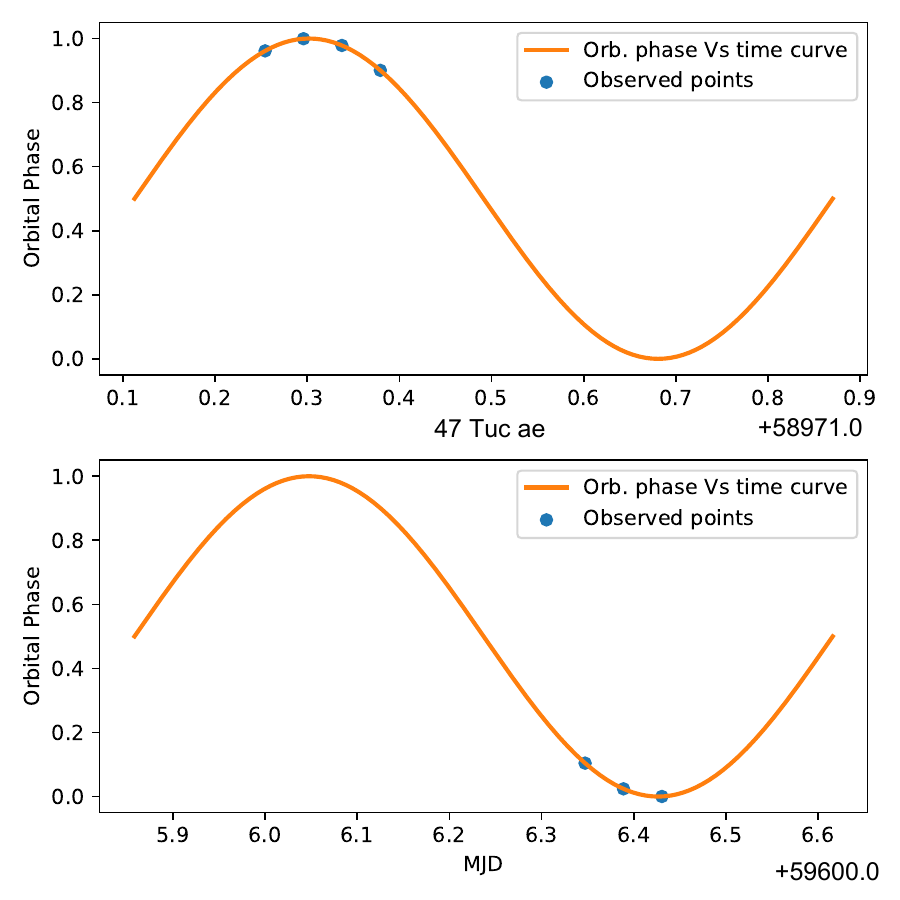}
    \caption{Orbital phase vs MJD plot for the binary pulsar 47~Tuc~ae from observations 10L (top) and 26U1 (bottom). The orbital phase of 0.5 corresponds to periastron passage. The orbital parameters are listed in Table \ref{tab:discovery_list}.}
    \label{fig:47Tuc_ae_10L_26U_orbphase_vs_MJD}
\end{figure}

47~Tuc~ae was discovered in or near the central beams of observations from both campaigns. It has a spin period of 3.87 ms and a DM of 24.24 pc cm$^{-3}$. The preliminary orbital calculation using the period-acceleration diagram (\citealt{2001MNRAS.322..885Freire}) showed an orbital period of approximately 18 hours. The orbital parameters were significantly improved by fitting Time of Arrivals (ToAs) with \textsc{tempo}, while including the 'JUMPs' between the ToA groups. Figure \ref{fig:47Tuc_ae_10L_26U_orbphase_vs_MJD} shows the orbital phase versus Modified Julian Dates (MJD) 
for the two brightest detections of this pulsar. The lower mass limit of the companion of the mass function is $0.121 \, \rm M_{\odot}$. No eclipses have been detected, suggesting that this is an MSP - He WD system. 

\subsubsection{47~Tuc~af}
\begin{figure}
    \centering
    \includegraphics[width=0.48\textwidth]{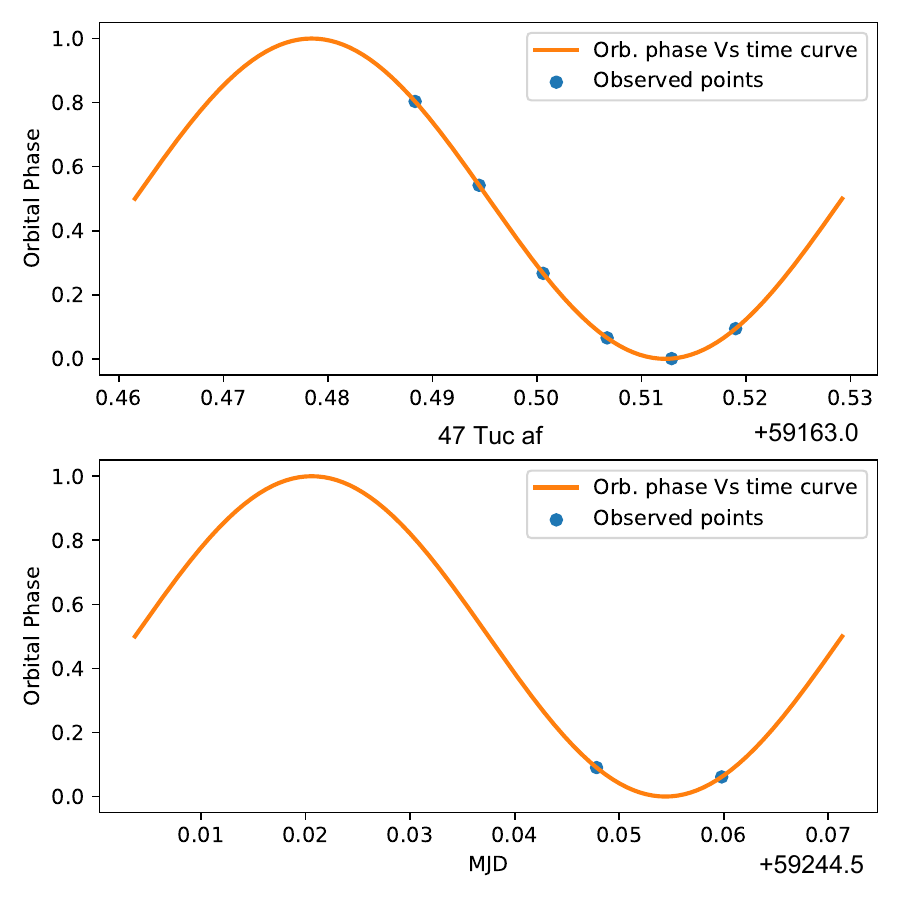}
    \caption{Orbital phase vs MJD plot for the binary pulsar 47~Tuc~af from observations 12U (top) and 19U (bottom). The orbital phase of 0.5 corresponds to periastron passage. The orbital parameters are listed in Table \ref{tab:discovery_list}. }
    \label{fig:47Tuc_af_12U_19U_orbphase_vs_MJD}
\end{figure}

47~Tuc~af was discovered in the central beams of some of the observations from both campaigns. It has a spin period of 2.99 ms with a DM of 24.34 pc cm$^3$. The detections of this pulsar are mostly in the UHF band, and only one detection is in the L band. Using the nine detections in total, we measured ToAs and fitted them with JUMPs (similar to 47~Tuc~ae in Sect. \ref{sec:47Tuc_ae}) and determined the accurate orbital parameters. Figure \ref{fig:47Tuc_af_12U_19U_orbphase_vs_MJD} shows the orbital phase as a function of MJD. The orbital period is very short (1.62 h), comparable to the slightly shorter orbital period of 47~Tuc~R (1.49 h) and one of the shortest among all binary pulsars in 47~Tuc. The lower limit on the mass of the companion is $0.0679\, \rm M_{\odot}$.  This mass is somewhat intermediate between that of BWs and RBs. The MeerKAT observation 10L and a detection with Parkes radio telescope's archival data (see Sect. \ref{sec:parkes_redetections}) were long enough to cover several orbits of the pulsar, in which the eclipses were seen, which indicates that it is clearly an 'eclipsing spider' system. 

\begin{figure}
    \centering
    \includegraphics[width=0.2\textwidth]{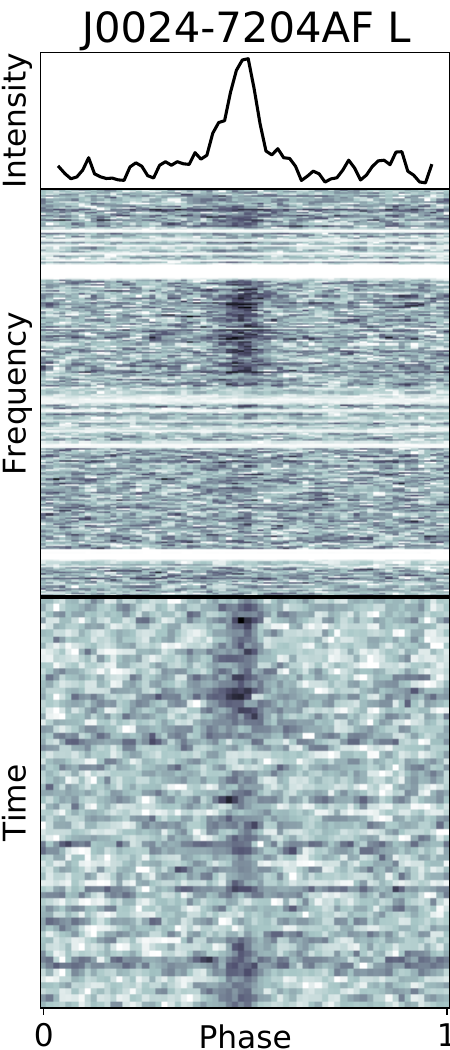}
    \includegraphics[width=0.2\textwidth]{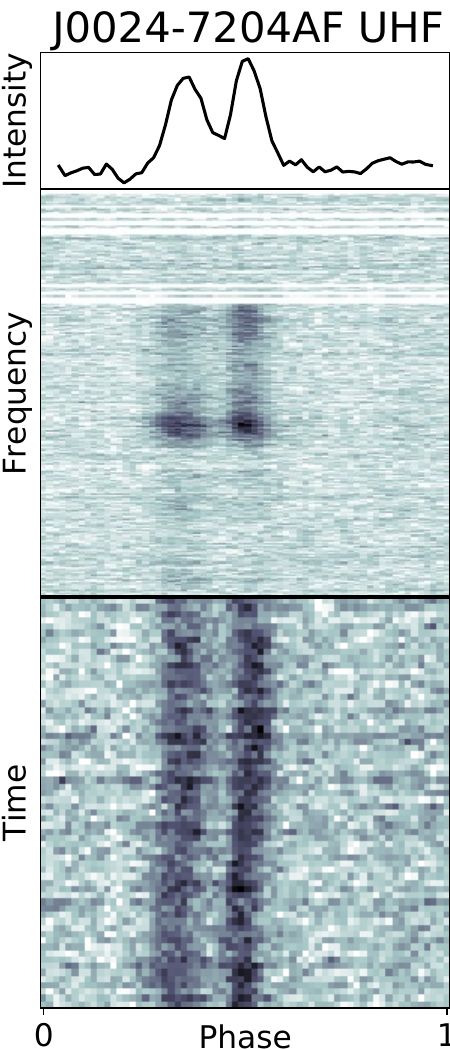}
    \caption{Pulse profiles of 47~Tuc~af folded using an L-band observation in the left and an UHF-band observation in the right. The former lasted approximately four hours, and the latter lasted approximately one hour.}
    \label{fig:J0024$-$7204AFi_profiles}
\end{figure}

Furthermore, the pulse profile in the UHF band clearly exhibits two components, whereas in the L-band no distinct multi-component structure can be seen. These differences can be seen in Fig. \ref{fig:J0024$-$7204AFi_profiles}.  Its nature is discussed in more detail in Sect. \ref{47tuc_af_discussion}.
 
\subsubsection{47~Tuc~ag}
47~Tuc~ag was discovered near the centre of the cluster and in only two observations separated by more than one year. It has a spin period of 9.76 ms and a DM of 24.5 pc cm$^{-3}$. This makes it the second slowest in the pulsar population of 47~Tuc.  From its two detections with MeerKAT (11L and 26U5 in Tables \ref{tab:observation_list_1}  
and \ref{tab:observation_list_2}) and a re-detection in the archival data of the Parkes radio telescope  (Sect.~\ref{sec:parkes_redetections} and Fig.~\ref{fig:redetection_ag}), we estimated the orbital parameters and listed them in Table~\ref{tab:discovery_list}. The orbital parameters imply a minimum companion mass of $\approx$ 0.01 M$_\odot$. More detections are necessary to allow a better determination of the orbital parameters.

\subsubsection{47~Tuc~ah}
47~Tuc~ah was also discovered in only two observations near the centre of the cluster. It has a spin period of 3.07 ms and a DM of 24.37 pc cm$^{-3}$. A significant period derivative indicates that it is in a binary system. The orbit of this pulsar is not well-constrained; a more accurate solution can only be derived with more detections in the future.   

\subsubsection{47~Tuc~ai}
47~Tuc~ai was discovered in an L-band observation during the first campaign and was subsequently detected in several UHF observations in the second campaign near the centre of the cluster. Its spin period of 13.03 ms makes this the slowest pulsar known in 47~Tuc. As shown in Fig.~\ref{fig:profiles_of_new_pulsars}, the pulse profile shows an additional component in the UHF band.

The spin period varies significantly between observations, which indicates that it is in a binary system. From the mass function, we infer that it has the largest minimum companion mass for any binary system in 47~Tuc. Although it was detected in multiple observations, deriving an orbital solution was challenging due to its eccentric ($e = 0.18$) orbit. Additional details on the follow-up of this system are provided by \cite{Risbud2026}.

\subsubsection{47~Tuc~ak}
47~Tuc~ak was discovered in 21U in the outer region of the cluster. It has a spin period of 3.52 ms and a DM of 23.91 pc cm$^{-3}$. It is one of only three pulsars in this cluster that is not near the cluster core (see inset in Fig.~\ref{fig:15L_tiling}), the other two being 47~Tuc~P (see below) and 47~Tuc~X \citep{Ridolfi2016}. The single detection in observation 21U shows a significant apparent spin period derivative and jerk over the observation of one hour, and hence confirms its binary nature.  

Due to its large offset from the cluster centre, the CB tilings of the second campaign did not cover its position. Hence, with only one detection, the orbital parameters could not be determined. Future observations will be made with its localisation position covered. 

\subsubsection{47~Tuc~al}
\begin{figure}
    \centering
    \includegraphics[width=0.48\textwidth]{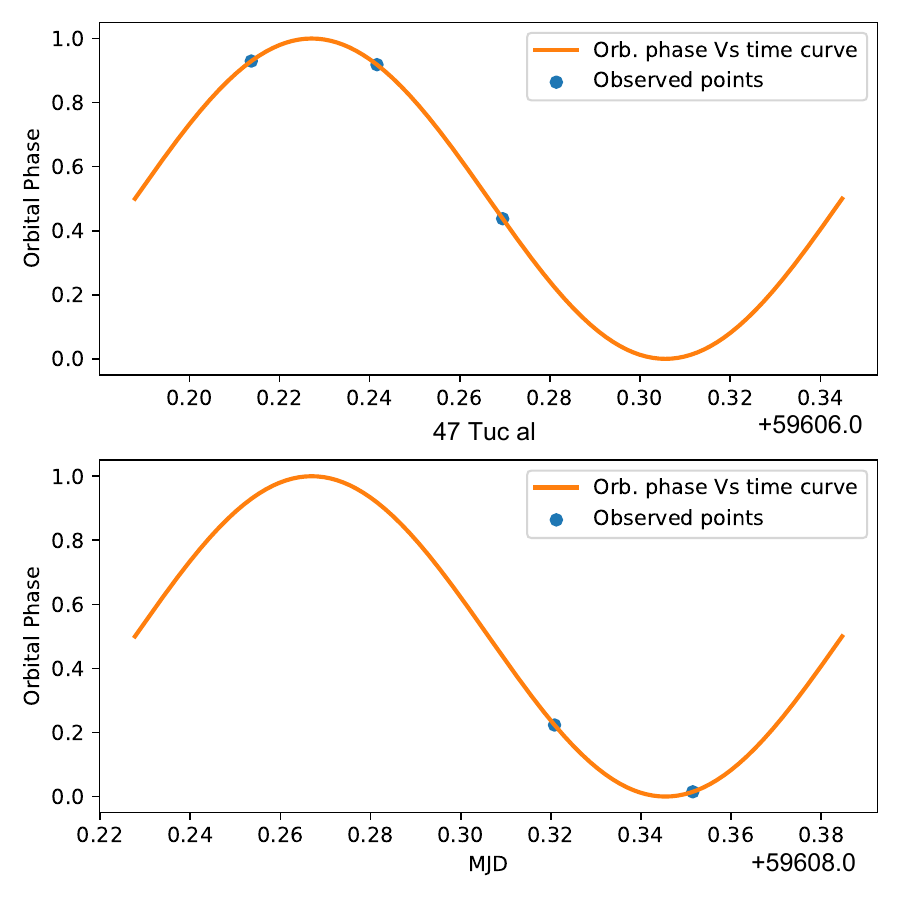}
    \caption{Orbital phase vs MJD plot for the binary pulsar 47~Tuc~al from observations 26U1 (top) and 28U (bottom). The orbital phase of 0.5 corresponds to periastron passage. The orbital parameters are listed in Table \ref{tab:discovery_list}. }
    \label{fig:47Tuc_al_orbphase_vs_MJD}
\end{figure}

47~Tuc~al was discovered slightly offset from the core in numerous observations in both campaigns. It has a spin period of 2.67 ms and a DM of 24.12 pc cm$^{-3}$. A faint interpulse can be seen in the integrated pulse profile shown in Fig.~\ref{fig:profiles_of_new_pulsars}. A total of 12 detections provide good coverage of its orbital phase, from which we derived an accurate orbital solution by fitting the ToAs. The system has an orbital period of 3.7 hours and the lower limit on its companion mass is $0.0091\, \rm M_{\odot}$, indicating that it is a BW system with an unusually low-mass companion similar to that of 47~Tuc~ac. The orbital period versus MJD diagram is shown in Fig. \ref{fig:47Tuc_al_orbphase_vs_MJD}. 

\subsubsection{47~Tuc~am}
47~Tuc~am was detected in only two adjoined epochs in the second campaign. It has a spin period of 4.16 ms and a DM of 24.50 pc cm$^{-3}$. It was localised with multi-beam detection, but the error on the position is relatively large because the pulsar was detected in the edge of the CB tiling and thus the pulsar was only partially surrounded by CBs. There is a difference in the spin period between the two observations, albeit small. Furthermore, the folding parameters for the second observation were not consistent with those from the first observation; therefore, we consider it to be in a binary system provisionally.

\subsubsection{47~Tuc~an, aq, ar, and as}
These pulsars were each detected in a single observation, and no additional conclusive detections were found in either campaign. Their observed properties are listed in Table \ref{tab:discovery_list}. Due to the limited number of detections, the orbital parameters remain undetermined. They are considered discoveries based on their detection in multiple surrounding beams (from which their localisation was obtained) and also because their signals show apparent drifting in the time versus phase diagram, indicating that their periods are changing through time in a physical way. Because of the low S/N of their detections, their positions have large uncertainties.

\subsection{Detection of previously known pulsars}

The sensitivity of the TRAPUM globular cluster surveys not only allows us to detect weak pulsars, which were rarely detected in the past, but it also gives immediate localisation to within a few arc-seconds with the help of multiple-beam localisation techniques. In these two campaigns, we managed to localise all previously known pulsars that lacked precise positions; these coordinates are listed in Table~\ref{tab:localized_position_known_pulsar}. 

\subsubsection{47~Tuc~P}

47~Tuc~P is in a 3.5-h binary system, originally discovered by \cite{Camilo2000} with the Murriyang 64-m (Parkes) radio telescope. They obtained a full coverage of the orbit from which the orbital solution was derived. This work was later extended by \cite{Ridolfi2016}, who obtained four additional detections with the same telescope. With that, its orbital parameters were improved through timing analysis over a time span of approximately 9.8 years. However, the lack of a phase-connected solution precluded a precise localisation.  
In the TRAPUM observations, it was detected in multiple observations during the first campaign. Its position was quickly narrowed down to the CB that has the brightest detection. The multiple-beam detections allowed for an improved localisation with \textsc{SeeKAT}. It is another of the three pulsars in 47~Tuc with a significant offset from the cluster centre.

\subsubsection{47~Tuc~V}
\label{chap:J0024$-$7204V}
47~Tuc~V was also first detected by \cite{Camilo2000} and later searched by \cite{Ridolfi2016}. The latter work yielded multiple additional detections, which allowed a precise determination of its orbital parameters; this and the presence of eclipses led to its classification as a redback system. In addition to regular eclipses during orbital conjunction, there are also irregular eclipses, i.e. short, well-defined eclipses that happen at all orbital phases. Furthermore, there are periods of a few years during which the pulsar is not detectable in any observation, despite the use of the \textsc{spider\_twister} routine\footnote{The use of this routine, which searches in orbital phase, is necessary because, as in most redback systems, the orbital period has irregular and unpredictable variations, which make it impossible to recover the pulsar signal just by folding the data with the pulsar's ephemeris.}.

During the TRAPUM campaigns, we re-detected this pulsar in observation 10L using \textsc{spider\_twister}. It was only visible for approximately one hour and only detected in one other observation. As mentioned above, this allows us to localise the pulsar within the size of the beam in which it was detected, and it is further localised with the detections from surrounded beams. Intriguingly, the localised position of this pulsar is close to a pulsar candidate discovered through MeerKAT imaging reported in \cite{Heywood2023}; this is discussed in more detail in Sect.~\ref{sec:V}.

\subsubsection{47~Tuc~ac}
47~Tuc~ac is a 2.74-ms pulsar in an eclipsing black widow system with an orbital period of 3.6 h originally discovered by \cite{Ridolfi2021}. It was detected in several observations in this work, albeit relatively weakly. The DM of the pulsar varied significantly between observations, from 24.6 to 25.0 pc cm$^{-3}$. In all cases, the signal was detectable only in segments shorter than one hour. This is likely due to a combination of interstellar scintillation and eclipses due to intra-binary materials, as also observed by \cite{Ridolfi2021}. Their observations on 47~Tuc did not utilise the FBFUSE backend to form multiple beams, hence the position of the pulsar was only constrained within the CB of 0.5 arcminute. In the TRAPUM observations, the position was narrowed down further as a result of the smaller CB size and multiple-beam detections. In addition, the new detections should provide a more complete coverage of the orbital phase to determine the orbital solution.

\subsubsection{47~Tuc~ad}
47~Tuc~ad is a 3.74-ms in a redback system with an orbital period of 7.64 h. It was detected in both campaigns. In one observation, the signal was bright for the final half hour of the two-hour observation. However, for the rest of the observation, it was barely visible after careful optimisation of the folding parameters. Similar behaviour was also presented in \cite{Ridolfi2021}, who originally discovered the pulsar. They reported that the signal appeared bright for only 1.28 hours of a seven-hour observation and disappeared for the rest of the time. Similarly to 47~Tuc~ac, its position was undetermined then. In the TRAPUM observations presented here, not only was the position constrained within the size of a single CB, but it was further localised using \textsc{SeeKAT} based on the detection in the surrounding beams.    

\begin{table}[h]
    \centering
    \caption{Localisation results of the known pulsars that were previously unlocalised.}
    \begin{tabular}{lll}
        \hline \hline
        Pulsar & \multicolumn{1}{c}{$\alpha$} & \multicolumn{1}{c}{$\delta$}  \\
        \hline
        P & 00:24:29(1) & $-$72:02:59(1)  \\
        V & 00:24:02(11) & $-$72:05:37(90)$^{\sharp}$  \\
        ac& 00:24:05(3) & $-$72:04:52(24)$^{\ddag}$ \\
        ad& 00:24:04(1) & $-$72:05:04(2) \\
        \hline
    \end{tabular}
\tablefoot{The symbol $\ddag$ indicates that the quoted position corresponds to the central coordinates of the coherent beam in which the pulsar was detected. The symbol $\sharp$ denotes cases where the large errors arise from multiple widely separated probability islands occurring in the \textsc{SeeKAT} localisation. The actual errors are shown in Fig. \ref{fig:V}.}
    \label{tab:localized_position_known_pulsar}
\end{table}

\section{Re-detections with Murriyang 64-m (Parkes) archival data }\label{sec:parkes_redetections}

The Murriyang 64-m (Parkes) radio telescope has been observing the GC 47~Tuc since 1990, resulting in the discovery and timing of many pulsars (see Sect. \ref{sec:intro} and references therein). The telescope carried out regular observations of 47~Tuc from 1997 to 2013, primarily with the multi-beam receiver \citep{Staveley-Smith1996}. This archival dataset includes 519 pointings\footnote{The logs of all 519 pointings can be found at \url{https://www3.mpifr-bonn.mpg.de/staff/pfreire/47Tuc/Observations_Table.html}.  } split across 414 observing epochs (i.e. days) that total around 1770 observing hours. The dataset and previous applications were described in \citet{Ridolfi2016}. We used this dataset in multiple ways to re-detect the newly discovered pulsars.

For the binary pulsars 47~Tuc~ae, af, and al, the orbits were relatively well-constrained such that their ephemerides could accurately fold the MeerKAT observations. Hence, for these pulsars, we used the archival Parkes observations, which were de-dispersed at the respective DMs and folded with \textsc{spider\_twister} to search for signals of the pulsar, which are summarised in Table \ref{tab:parkes_redetections}; 47~Tuc~al is the most remarkable case with 13 re-detections. The brightest of these re-detections are shown in Figs. \ref{fig:redetection_ag} to \ref{fig:redetection_al}.  

The binary pulsars 47~Tuc~ag, ah and ak were only detected sporadically in the MeerKAT observations, and hence their orbital parameters could not be constrained. For these pulsars, we applied the \textsc{spider\_twister} search as before, and in addition, we also performed blind acceleration searches at their respective DMs. As expected, this resulted in very few detections (one of 47~Tuc~ag and two for 47~Tuc~ak)  summarised in Table \ref{tab:parkes_redetections}. These may eventually turn out to be important in constraining the orbit parameters in follow-up works. 47~Tuc~ae and ah were not detected by any method; on top of the sensitivity limitation of the Parkes radio telescope, this non-detection may be due to their intrinsic faintness, unfavourable scintillation, and because their orbital parameters are not yet good enough for \textsc{spider\_twister} folding. 

Pulsars 47~Tuc~am to 47~Tuc as had extremely faint detections even with the MeerKAT, and their orbital parameters could not be determined. Hence, for these systems, we performed segmented acceleration searches at their respective DMs, using the full length and one-hour observation segments.  Given the faintness of their detections with MeerKAT, these pulsars would be undetectable with the less sensitive Parkes data for any integration length smaller than one hour, so smaller timeseries segments were not searched. No detections of any of these were found.

\begin{table}[h]
    \centering
    \caption{Re-detections of newly discovered pulsars in the archival Parkes data (see Sect. \ref{sec:parkes_redetections}).}
    \begin{tabular}{lll}
        \hline \hline
        Pulsar & No. of Detections & $\sigma_{best}$   \\
        \hline
        af & 2 & 11.1  \\
        ag & 1 & 5.6   \\
        ai & 3 & 5.8 \\
        ak & 2 & 9.4 \\
        al & 13 & 21.2  \\
        \hline
    \end{tabular}
\tablefoot{The second column gives the total number of re-detections over the Parkes datasets span from 1997 to 2013, and the third column gives the \textsc{PRESTO} detection significance ($\sigma$) of the brightest re-detection. Newly discovered pulsars that are not listed in this table were not detected in the Parkes data.}
    \label{tab:parkes_redetections}
\end{table}

\section{Discussion}
\label{sec:discussion}
These 15 new discoveries add to the two previous MeerKAT discoveries, marking the first significant increase in the number of pulsars known in 47~Tuc since the observations with the Parkes multi-beam system in the late 1990s and early 2000s \citep{Camilo2000}. The total pulsar population of the cluster now stands at a total of 42. These observations also resulted in consistent detections of previously known pulsars and frequent detections of elusive systems such as P, V \citep{Ridolfi2016}, ac, and ad \citep{Ridolfi2021}, which were never detected or only rarely, with the Parkes radio telescope. This is an illustration of the immediate effect of the increased sensitivity provided by MeerKAT.

It should be noted that, as for the Parkes detections, the large scintillation amplitudes observed for the pulsars in this globular cluster mean that new pulsars continue to be discovered when scintillation happens to bring their flux density above the sensitivity threshold of the telescope. In this regard, the number of MeerKAT observations of 47~Tuc is still small compared to the number of observations reported by \cite{Camilo2000}. This means that as timing observations proceed, the number of MeerKAT discoveries will likely increase; the same will happen with the number of detections of already discovered pulsars for which the orbits are still undetermined.

Of the 27 previously known pulsars in 47~Tuc, 17
($\approx$ 63 \%) were in binaries; with the discoveries in this work, the binary fraction increases to 29 / 42 $\simeq$ 69\%.
Only four of the new 12 binaries have well-determined orbits.

Most of the known binary systems in 47~Tuc have low eccentricities and the pulsars have low-mass companions, with the exception of 47~Tuc
~H, which has an eccentricity of 0.07 and to a much lesser extent 47~Tuc~E~and~Tuc S, which have orbital eccentricities of 0.0003.
However, one of the new discoveries in this work, 47~Tuc~ai, is unusual because of its large companion mass and orbital eccentricity of 0.18. As mentioned above, this system is discussed in detail by \cite{Risbud2026}.
 
Although relatively small, these eccentricities are orders of magnitude larger than those of similar systems in the Galaxy; this is likely a consequence of encounters with other stars in the GC \citep[see][for a review]{Phinney1992}. Nevertheless, these orbital eccentricities are significantly lower than those observed in other GCs such as Terzan 5 \citep{Ransom2005,Andersen2018,Padmanabh2024} and NGC 1851 \citep{Freire2004,Ridolfi2022}.

The longest orbital period of a binary pulsar in 47~Tuc is that of 47~Tuc~X, $\sim$11 d \citep{Ridolfi2016}. However, this system is located relatively far from the cluster centre ($\sim$3.8 arcminutes) and its very low orbital eccentricity shows that it has been little perturbed by close interactions with other stars in the cluster. Apart from this system, all other binaries with well-determined orbits have $P_{\rm b} < 2.3\, \rm d$ (for 47~Tuc~H). The four new pulsars with  well-determined orbits confirm, thus far, the trend of short orbital periods.

The stellar encounter rate of 47~Tuc is \textbf{$\Gamma$} $\approx$ 1000 \citep{Bahramian2013}, which, although it is the second highest in Galactic GCs after Terzan 5, is still a factor of seven lower than that of the latter cluster. The main difference from Terzan 5 is the much lower value of this encounter rate divided by the number of stars $\gamma$, which suggests that once formed, new LMXBs are much less likely to be perturbed by subsequent exchange encounters. This could explain the much smaller number of MSPs with eccentric orbits and massive companions in 47~Tuc compared to Terzan 5 or NGC 1851, where such highly perturbed systems are abundant \citep{Ransom2005,Andersen2018,Padmanabh2024,Freire2004,Ridolfi2022,Barr2024}.

However, it is interesting to remark that in NGC~1851 and especially Terzan 5 there are many wider binary systems, with orbital periods up to 60 days, and the existence and survival of such systems in a GC with a much larger $\gamma$ than that of 47~Tuc might seem surprising at first. Perhaps this is caused by the evolutionary stages of the GCs: in Terzan 5, these systems are forming and are being destroyed; in NGC~1851 even more eccentric binaries are formed, with many eccentric binaries being disrupted or becoming tighter; in 47~Tuc the MSP population might reflect a more distant past where the wider, eccentric systems have long been destroyed (presumably in an earlier core collapse phase of the cluster evolution), and only tighter, more circular systems remain in the core-rebound phase. The population of binaries in 47~Tuc is especially reminiscent of another very dense GC, M62: in the latter GC, all ten known pulsars are in binaries, and among the nine with known orbits, eight have orbital periods shorter than 1.1 d, all in low-eccentricity orbits with low-mass companions \citep{Vleeschower2024}.

\subsection{Optical identification of 47~Tuc~af?}
\label{47tuc_af_discussion}

The discovery of 47~Tuc~W, a typical redback system \citep{Camilo2000}, led to the identification, in HST optical data, of a tidally locked companion (W29$_{\rm opt}$, so named because it coincides with the X-ray source W29, \citealt{Grindlay2001}) with a large temperature difference between the hemisphere that is irradiated by the pulsar and the opposite hemisphere. 
The resulting variation in the optical magnitude of this star with time allowed a precise measurement of its orbital period, which coincides precisely with the orbital period of 47~Tuc~W \citep{Edmonds2002}. The timing solution for this pulsar \citep{Ridolfi2016} later confirmed its association with W29$_{\rm opt}$.

In addition to 47~Tuc~W, \cite{Edmonds2002} identified a second candidate binary companion of a (then undetected) MSP, W34$_{\rm opt}$. Its position is $\alpha =$ 00h 24m 05.21(2)s, $\delta = -72^\circ$ 04\arcmin 46\farcs59(6), which makes it virtually coincident with the X-ray source W34.
The characteristics of W34$_{\rm opt}$ are very similar to the optical characteristics of the companion of 47~Tuc - W: the overall I and V magnitudes of the two systems are identical; however, W34$_{\rm opt}$ is significantly brighter in the U band (22.3 vs 23.8 for W29$_{\rm opt}$). The amplitude of the modulation is also similar, but 30\% larger than for W29$_{\rm opt}$. All of this suggests a star that is very similar to the redback companion of 47~Tuc~W but with an even more dramatic difference in brightness and temperature between the irradiated and non-irradiated hemispheres, as might be expected from the significantly more compact orbit.

The large amplitude variations of the optical magnitude allowed a precise measurement of
the orbital period, $P_{\rm b} = 0.0676705(78)\,$d.
This orbital period coincides, within uncertainties, with the orbital period of 47~Tuc~af, suggesting a possible association.

\begin{figure}[h]
    \centerline{
    \includegraphics[width=\columnwidth]{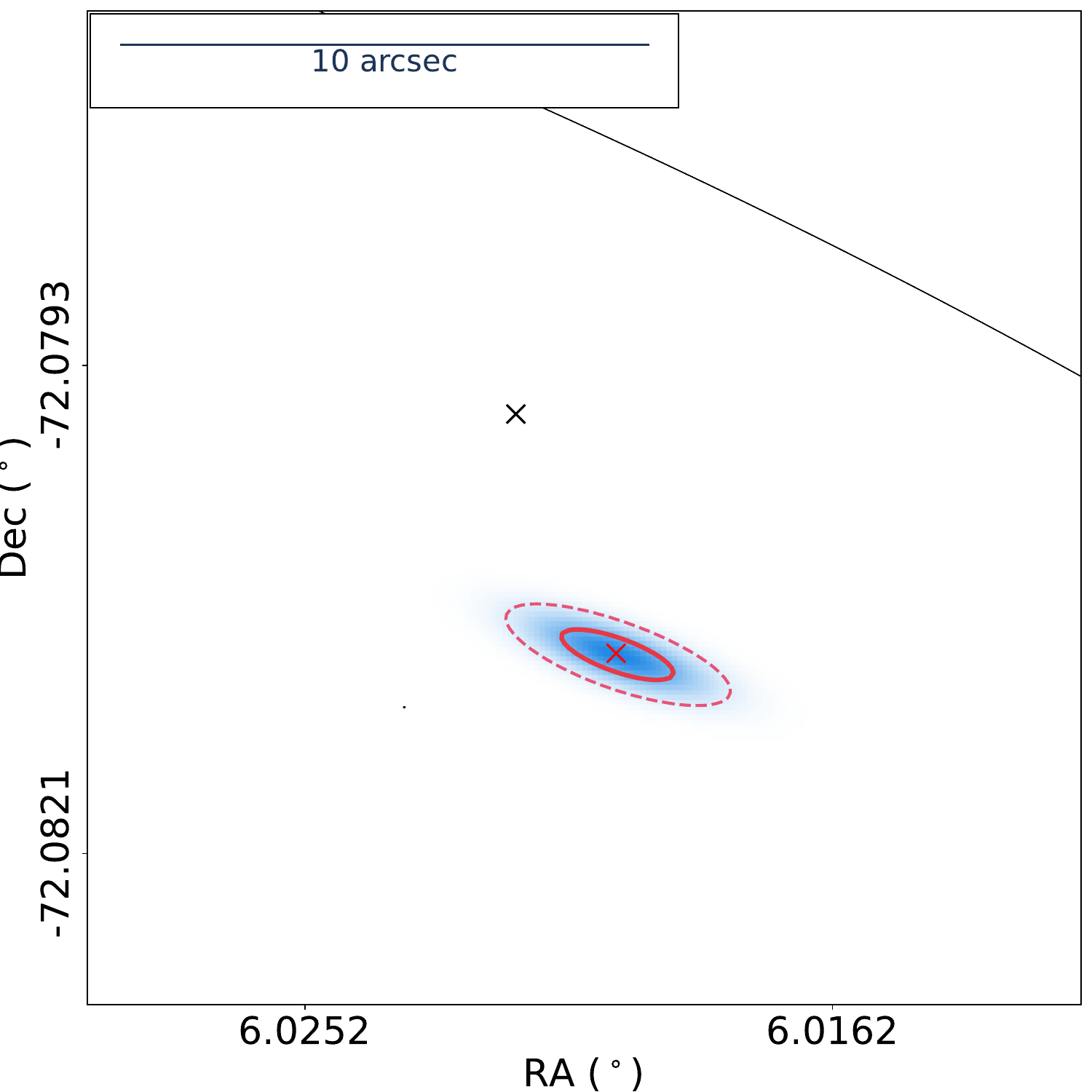}
    }
    \caption{Localisation of 47~Tuc~af. The red cross indicates the best position derived with \textsc{SeeKAT}, while the black cross indicates the position of W34$_{\rm opt}$. The solid red contour indicates 1$\sigma$ confident level and the dashed red contour indicates 2$\sigma$ confident level. The black line indicates the edge of the beam at the corresponding gain level.}
    \label{fig:af}
\end{figure}

However, as we can see in Fig.~\ref{fig:af}, the \textsc{SeeKAT} position of 47~Tuc~af listed in Table~\ref{tab:discovery_list},
shown in red in the figure, does not coincide with the position of W34$_{\rm opt}$, which is indicated with the black cross; indeed, the
latter lies well outside the position error circle of 47~Tuc~af. Intriguingly, the positional offset is only approximately 5 arcseconds.
This begs the question of why two binaries with exactly the same orbital periods should be located so close to each other.

Due to this mismatch, the identification of 47~Tuc~af with W34$_{\rm opt}$, which is based on the perfect match of orbital periods, remains to be confirmed. A future phase-connected timing solution for this pulsar will clarify the matter and confirm or exclude this identification.

There is a final interesting fact about the companion of 47~Tuc~af, its mass: the system's mass function implies a minimum companion mass of 0.068 $\rm M_{\odot}$, which is little more than half the minimum companion mass of 47~Tuc~W, $0.127 \, \rm M_{\odot}$. The mass of the companion of 47~Tuc~af is somewhat intermediate between the companion masses of black widows and redbacks. This could be due to a lower orbital inclination; however, a low orbital inclination is not consistent with the large amplitude of the variation of the optical magnitude as a function of orbital phase. A lower companion mass might be expected for such compact systems, where the smaller size of the Roche lobe and stronger irradiation from the pulsar lead to accelerated mass loss.

\subsection{Is 47~Tuc~V the source seen by Heywood et al?}

\label{sec:V}
\begin{figure}[h]
    \centering
    \includegraphics[width=\columnwidth]{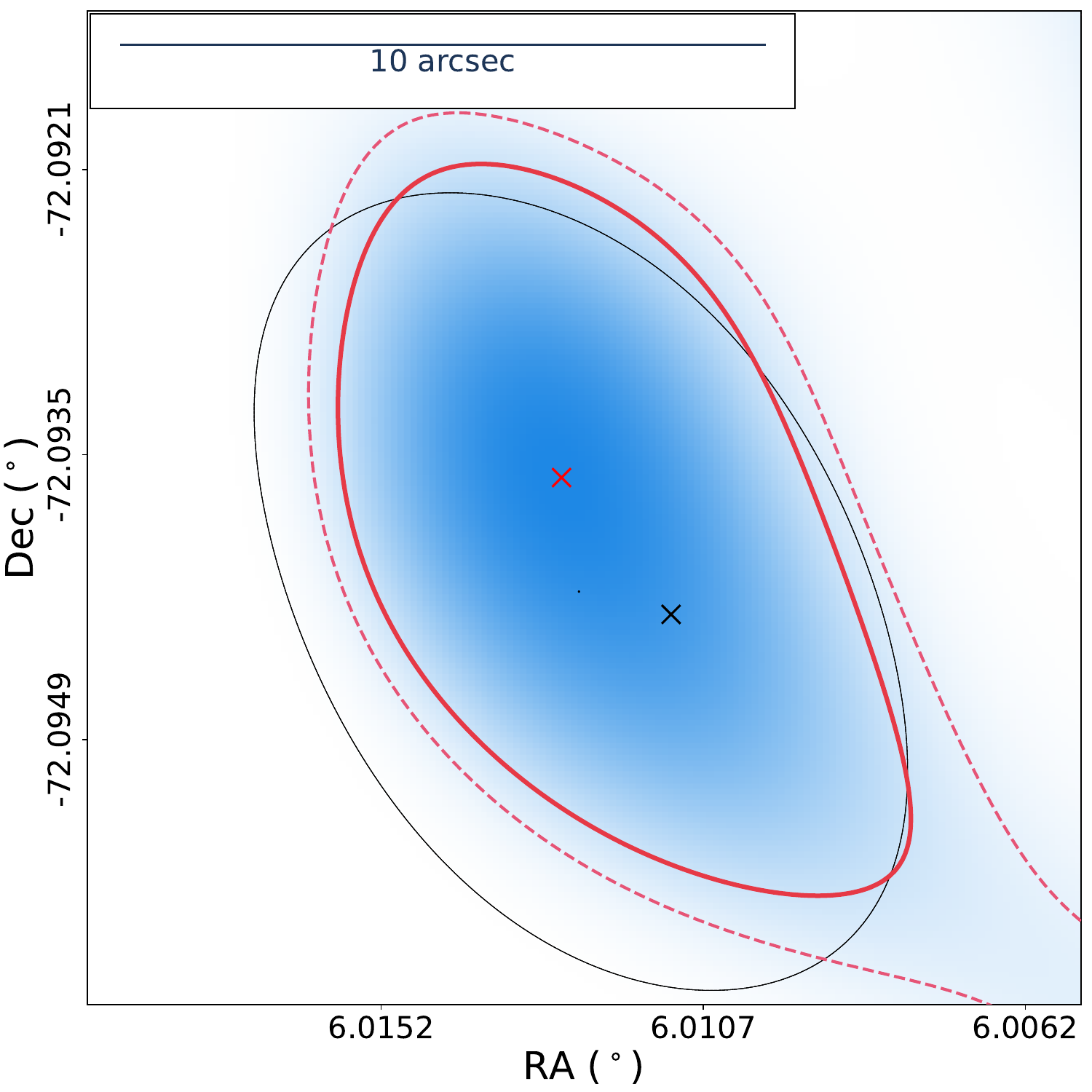}
    \caption{\textsc{SeeKAT} localization of 47~Tuc~V. The red cross indicates the best position given by \textsc{SeeKAT} and the red contour its estimated uncertainty, while the black cross indicates the position given by \cite{Heywood2023}. The positions of both sources match very well within their relative uncertainties. Please refer to Fig.~\ref{fig:af} for the meaning of the lines.}
    \label{fig:V}
\end{figure}

The radio images of 47~Tuc produced by \cite{Heywood2023} include most known pulsars with timing solutions, but intriguingly they also reveal an additional source, J002402.7$-$720539.4, which was identified as a candidate MSP. This identification was made based on its variability, which has characteristics (amplitude, timescale) similar to those of the remaining MSPs. For most of these MSPs, this variation comes from diffractive scintillation, but for a few, additional changes in flux are caused by eclipses. This is also the case for 47~Tuc~V, where the eclipses occur at all orbital phases. Given the similarities with the other MSPs, \cite{Heywood2023} suggested that this could be one of the few pulsars for which no positions were known, including 47~Tuc~V.

In Fig.~\ref{fig:V} we can see that the \textsc{SeeKAT} position derived from our detections of 47~Tuc~V coincides well within uncertainties with the position of J002402.7$-$720539.4 derived by \cite{Heywood2023}, strongly suggesting that they are the same source.

\section{Summary and future prospects}
\label{sec:summary}
In this paper, we report the discoveries of 15 pulsars from our TRAPUM campaigns on 47~Tucanae, which add to two previous TRAPUM discoveries, 47~Tuc~ac and ad \citep{Ridolfi2021}, for a total of 17 TRAPUM + MeerTime / MeerKAT discoveries; this raises the number of known pulsars in 47~Tuc to 42. The new discoveries include three isolated pulsars and 12 binary pulsars, which increase the total binary fraction in the cluster to 69\%. The new binary pulsars with well-determined orbits have characteristics that are similar to those of the previously known pulsars: low mass companions and orbits with relatively small periods and eccentricities, the exception being 47~Tuc~ai, which will be discussed by \cite{Risbud2026}.

One of the new systems presented in this work, 47~Tuc~af, is likely associated with a bright optical MSP candidate detected in HST imagery in 2002. In addition to localising the new discoveries, we
also localised four previously known binary pulsars with no known timing solutions. One of these, 47~Tuc~V,  coincides spatially with the source reported by \cite{Heywood2023}.

Because of the strong scintillation screen through the line of sight, the detections of these new pulsars are highly sporadic. This implies that for some of the new discoveries, many additional observations will be necessary to fully characterise them, as they will only be detectable in a small number. For this reason we had to localise two pulsars discovered by \cite{Camilo2000}, 47~Tuc~P and V, as they still lack timing solutions.
However, one can look at the situation positively: each new observation of 47~Tuc has the potential for both new discoveries and new useful detections of previously known pulsars, which will help determine additional orbits.

Despite the sparsity of detections, the orbits of these elusive pulsars can be determined with the aforementioned period-acceleration method, which was first used to determine the orbits of 47~Tuc~S and T, and the subsequent use of the orbital phase-finding method described by \cite{Ridolfi2016}, as implemented in \textsc{spider\_twister}, which was first used to re-detect and determine precise orbits for 47~Tuc~P and V. Both methods have already been used on the determination of the orbits of some of the new pulsars in this work.
Even after a well-determined orbit, the ToAs can still be very extremely sparse. However, even such sparse ToA sets can sometimes be connected using \textsc{dracula} \citep{FreireRidolfi2018}, which was first used to determine the timing solution of 47~Tuc~aa (which had only 18 detections in more than 500 Parkes observations). The problem of strong scintillation will be further mitigated with the use of the UHF band, where the average size of the scintles, both in time and in frequency, is considerably smaller, leading to more regular detections of all pulsars. Additionally, there will be another campaign of observations in 2026 with a total observing time of 24 hours to follow up the newly discovered pulsars.  

Therefore, further observations will have the potential for additional discoveries, determining additional orbits, establishing timing solutions for the new pulsars, and improving those of the previously known pulsars. Apart from precise positions, which will allow orbital multi-wavelength follow-up, they will result in many more measurements of the proper motion and line-of-sight accelerations of the systems, which will place additional constraints on the gravitational field of the globular cluster and further probe its content of ionised gas.

\begin{acknowledgements}
TRAPUM observations used the FBFUSE and APSUSE computing clusters for data acquisition, storage and analysis. These clusters were funded and installed by the Max-Planck-Institut für Radioastronomie and the Max-Planck-Gesellschaft.
The MeerKAT telescope is operated by the South African Radio Astronomy Observatory, which is a facility of the National Research Foundation, an agency of the Department of Science and Innovation.
WC, DR, PCCF, AR, EB, MK, PVP acknowledge continuing valuable support from the Max-Planck Society.
FA acknowledges that part of the research activities described in this paper were carried out with the contribution of the NextGenerationEU funds within the National Recovery and Resilience Plan (PNRR), Mission 4 – Education and Research, Component 2 – From Research to Business (M4C2), Investment Line 3.1 – Strengthening and creation of Research Infrastructures, Project IR0000034 – ‘STILES -Strengthening the Italian Leadership in ELT and SKA’. V.V.K. acknowledges financial support from the ERC starting grant ‘COMPACT’ (Understanding gravity using a COMprehensive search for fast-spinning Pulsars And CompacT binaries, grant agreement no. 101078094).
\end{acknowledgements}

\bibliographystyle{aa}
\bibliography{references.bib}

@ARTICLE{Andersen2018,
       author = {{Andersen}, Bridget C. and {Ransom}, Scott M.},
        title = "{A Fourier Domain {\textquotedblleft}Jerk{\textquotedblright} Search for Binary Pulsars}",
      journal = {\apjl},
         year = 2018,
        month = aug,
       volume = {863},
       number = {1},
          eid = {L13},
        pages = {L13},
          doi = {10.3847/2041-8213/aad59f},
archivePrefix = {arXiv},
       eprint = {1807.07900},
 primaryClass = {astro-ph.HE},
       adsurl = {https://ui.adsabs.harvard.edu/abs/2018ApJ...863L..13A}
}

@ARTICLE{Phinney1992,
       author = {{Phinney}, E.~S.},
        title = "{Pulsars as Probes of Newtonian Dynamical Systems}",
      journal = {Philosophical Transactions of the Royal Society of London Series A},
         year = 1992,
        month = oct,
       volume = {341},
       number = {1660},
        pages = {39-75},
          doi = {10.1098/rsta.1992.0084},
       adsurl = {https://ui.adsabs.harvard.edu/abs/1992RSPTA.341...39P}
}

@ARTICLE{FreireRidolfi2018,
       author = {{Freire}, Paulo C.~C. and {Ridolfi}, Alessandro},
        title = "{An algorithm for determining the rotation count of pulsars}",
      journal = {\mnras},
         year = 2018,
        month = jun,
       volume = {476},
       number = {4},
        pages = {4794-4805},
          doi = {10.1093/mnras/sty524},
archivePrefix = {arXiv},
       eprint = {1802.07211},
 primaryClass = {astro-ph.IM},
       adsurl = {https://ui.adsabs.harvard.edu/abs/2018MNRAS.476.4794F}
}

@ARTICLE{2012AJ....143...50Woodley,
       author = {{Woodley}, K.~A. and {Goldsbury}, R. and {Kalirai}, J.~S. and {Richer}, H.~B. and {Tremblay}, P. -E. and {Anderson}, J. and {Bergeron}, P. and {Dotter}, A. and {Esteves}, L. and {Fahlman}, G.~G. and {Hansen}, B.~M.~S. and {Heyl}, J. and {Hurley}, J. and {Rich}, R.~M. and {Shara}, M.~M. and {Stetson}, P.~B.},
        title = "{The Spectral Energy Distributions of White Dwarfs in 47 Tucanae: The Distance to the Cluster}",
      journal = {\aj},
         year = 2012,
        month = feb,
       volume = {143},
       number = {2},
          eid = {50},
        pages = {50},
          doi = {10.1088/0004-6256/143/2/50},
archivePrefix = {arXiv},
       eprint = {1112.1425},
 primaryClass = {astro-ph.GA},
       adsurl = {https://ui.adsabs.harvard.edu/abs/2012AJ....143...50W}
}

@ARTICLE{Staveley-Smith1996,
       author = {{Staveley-Smith}, L. and {Wilson}, W.~E. and {Bird}, T.~S. and
         {Disney}, M.~J. and {Ekers}, R.~D. and {Freeman}, K.~C. and
         {Haynes}, R.~F. and {Sinclair}, M.~W. and {Vaile}, R.~A. and
         {Webster}, R.~L. and {Wright}, A.~E.},
        title = "{The Parkes 21 CM multibeam receiver}",
      journal = {\pasa},
         year = 1996,
        month = nov,
       volume = {13},
       number = {3},
        pages = {243-248},
       adsurl = {https://ui.adsabs.harvard.edu/abs/1996PASA...13..243S}
}

@ARTICLE{2001MNRAS.322..885Freire,
       author = {{Freire}, P.~C. and {Kramer}, M. and {Lyne}, A.~G.},
        title = "{Determination of the orbital parameters of binary pulsars}",
      journal = {\mnras},
         year = 2001,
        month = apr,
       volume = {322},
       number = {4},
        pages = {885-890},
          doi = {10.1046/j.1365-8711.2001.04200.x},
archivePrefix = {arXiv},
       eprint = {astro-ph/0010463},
 primaryClass = {astro-ph},
       adsurl = {https://ui.adsabs.harvard.edu/abs/2001MNRAS.322..885F}
}

@INPROCEEDINGS{Barr2018,
       author = {{Barr}, Ewan D.},
        title = "{An S-band Receiver and Backend System for MeerKAT}",
    booktitle = {Pulsar Astrophysics the Next Fifty Years},
       series = {IAU symposium and colloquium proceedings series},
         year = {2018},
       editor = {{Weltevrede}, P. and {Perera}, B.~B.~P. and {Preston}, L.~L. and {Sanidas}, S.},
       volume = {337},
        month = {08},
        pages = {175-178},
          doi = {10.1017/S1743921317009036},
       adsurl = {https://ui.adsabs.harvard.edu/abs/2018IAUS..337..175B}
}

@ARTICLE{Ridolfi2021,
       author = {{Ridolfi}, A. and {Gautam}, T. and {Freire}, P.~C.~C. and {Ransom}, S.~M. and {Buchner}, S.~J. and {Possenti}, A. and {Krishnan}, V. Venkatraman and {Bailes}, M. and {Kramer}, M. and {Stappers}, B.~W. and {Abbate}, F. and {Barr}, E.~D. and {Burgay}, M. and {Camilo}, F. and {Corongiu}, A. and {Jameson}, A. and {Padmanabh}, P.~V. and {Vleeschower}, L. and {Champion}, D.~J. and {Chen}, W. and {Geyer}, M. and {Karastergiou}, A. and {Karuppusamy}, R. and {Parthasarathy}, A. and {Reardon}, D.~J. and {Serylak}, M. and {Shannon}, R.~M. and {Spiewak}, R.},
        title = "{Eight new millisecond pulsars from the first MeerKAT globular cluster census}",
      journal = {\mnras},
         year = 2021,
        month = mar,
          doi = {10.1093/mnras/stab790},
archivePrefix = {arXiv},
       eprint = {2103.04800},
 primaryClass = {astro-ph.HE},
       adsurl = {https://ui.adsabs.harvard.edu/abs/2021MNRAS.tmp..783R}
}

@ARTICLE{Bailes2020,
       author = {{Bailes}, M. and {Jameson}, A. and {Abbate}, F. and {Barr}, E.~D. and {Bhat}, N.~D.~R. and {Bondonneau}, L. and {Burgay}, M. and {Buchner}, S.~J. and {Camilo}, F. and {Champion}, D.~J. and {Cognard}, I. and {Demorest}, P.~B. and {Freire}, P.~C.~C. and {Gautam}, T. and {Geyer}, M. and {Griessmeier}, J. -M. and {Guillemot}, L. and {Hu}, H. and {Jankowski}, F. and {Johnston}, S. and {Karastergiou}, A. and {Karuppusamy}, R. and {Kaur}, D. and {Keith}, M.~J. and {Kramer}, M. and {van Leeuwen}, J. and {Lower}, M.~E. and {Maan}, Y. and {McLaughlin}, M.~A. and {Meyers}, B.~W. and {Os{\l}owski}, S. and {Oswald}, L.~S. and {Parthasarathy}, A. and {Pennucci}, T. and {Posselt}, B. and {Possenti}, A. and {Ransom}, S.~M. and {Reardon}, D.~J. and {Ridolfi}, A. and {Schollar}, C.~T.~G. and {Serylak}, M. and {Shaifullah}, G. and {Shamohammadi}, M. and {Shannon}, R.~M. and {Sobey}, C. and {Song}, X. and {Spiewak}, R. and {Stairs}, I.~H. and {Stappers}, B.~W. and {van Straten}, W. and {Szary}, A. and {Theureau}, G. and {Venkatraman Krishnan}, V. and {Weltevrede}, P. and {Wex}, N. and {Abbott}, T.~D. and {Adams}, G.~B. and {Burger}, J.~P. and {Gamatham}, R.~R.~G. and {Gouws}, M. and {Horn}, D.~M. and {Hugo}, B. and {Joubert}, A.~F. and {Manley}, J.~R. and {McAlpine}, K. and {Passmoor}, S.~S. and {Peens-Hough}, A. and {Ramudzuli}, Z.~R. and {Rust}, A. and {Salie}, S. and {Schwardt}, L.~C. and {Siebrits}, R. and {Van Tonder}, G. and {Van Tonder}, V. and {Welz}, M.~G.},
        title = "{The MeerKAT telescope as a pulsar facility: System verification and early science results from MeerTime}",
      journal = {\pasa},
         year = 2020,
        month = jul,
       volume = {37},
          eid = {e028},
        pages = {e028},
          doi = {10.1017/pasa.2020.19},
archivePrefix = {arXiv},
       eprint = {2005.14366},
 primaryClass = {astro-ph.IM},
       adsurl = {https://ui.adsabs.harvard.edu/abs/2020PASA...37...28B}
}

@ARTICLE{Zhang2017,
       author = {{Zhang}, Fan},
        title = "{Pulsar magnetospheric convulsions induced by an external magnetic field}",
      journal = {\aap},
         year = 2017,
        month = feb,
       volume = {598},
          eid = {A88},
        pages = {A88},
          doi = {10.1051/0004-6361/201629254},
archivePrefix = {arXiv},
       eprint = {1701.01209},
 primaryClass = {astro-ph.HE},
       adsurl = {https://ui.adsabs.harvard.edu/abs/2017A&A...598A..88Z}
}

@INPROCEEDINGS{StappersKramer2016,
       author = {{Stappers}, B.~W. and {Kramer}, M.},
        title = "{An Update on TRAPUM}",
    booktitle = {MeerKAT Science: On the Pathway to the SKA},
         year = 2016,
        month = jan,
          eid = {9},
        pages = {9},
       adsurl = {https://ui.adsabs.harvard.edu/abs/2016mks..confE...9S}
}

@PHDTHESIS{Knight2007,
       author = {{Knight}, Haydon Stephen},
        title = "{Pulsar applications of baseband recording}",
       school = {Swinburne University of Technology},
         year = 2007,
        month = Jan,
    	   doi = {https://doi.org/10.25916/sut.26293105.v1}
}

@INPROCEEDINGS{Jonas2016,
       author = {{Jonas}, J. and {MeerKAT Team}},
        title = "{The MeerKAT Radio Telescope}",
    booktitle = {MeerKAT Science: On the Pathway to the SKA},
         year = 2016,
        month = jan,
          eid = {1},
        pages = {1},
       adsurl = {https://ui.adsabs.harvard.edu/abs/2016mks..confE...1J}
}

@ARTICLE{Chen2021,
       author = {{Chen}, Weiwei and {Barr}, Ewan and {Karuppusamy}, Ramesh and {Kramer}, Michael and {Stappers}, Benjamin},
        title = "{Wide Field Beamformed Observation with MeerKAT}",
      journal = {Journal of Astronomical Instrumentation},
         year = 2021,
        month = jan,
       volume = {10},
       number = {3},
          eid = {2150013-178},
        pages = {2150013-178},
          doi = {10.1142/S2251171721500136},
archivePrefix = {arXiv},
       eprint = {2110.01667},
 primaryClass = {astro-ph.IM},
       adsurl = {https://ui.adsabs.harvard.edu/abs/2021JAI....1050013C}
}

@ARTICLE{Ransom2003,
       author = {{Ransom}, Scott M. and {Cordes}, James M. and {Eikenberry}, Stephen S.},
        title = "{A New Search Technique for Short Orbital Period Binary Pulsars}",
      journal = {\apj},
         year = 2003,
        month = jun,
       volume = {589},
       number = {2},
        pages = {911-920},
          doi = {10.1086/374806},
archivePrefix = {arXiv},
       eprint = {astro-ph/0210010},
 primaryClass = {astro-ph},
       adsurl = {https://ui.adsabs.harvard.edu/abs/2003ApJ...589..911R}
}

@ARTICLE{Verbunt2014,
       author = {{Verbunt}, Frank and {Freire}, Paulo C.~C.},
        title = "{On the disruption of pulsar and X-ray binar ies in globular clusters}",
      journal = {\aap},
         year = 2014,
        month = jan,
       volume = {561},
          eid = {A11},
        pages = {A11},
          doi = {10.1051/0004-6361/201321177},
archivePrefix = {arXiv},
       eprint = {1310.4669},
 primaryClass = {astro-ph.SR},
       adsurl = {https://ui.adsabs.harvard.edu/abs/2014A&A...561A..11V}
}

@ARTICLE{Dewey1985,
       author = {{Dewey}, R.~J. and {Taylor}, J.~H. and {Weisberg}, J.~M. and {Stokes}, G.~H.},
        title = "{A search for low-luminosity pulsars.}",
      journal = {\apjl},
         year = 1985,
        month = jul,
       volume = {294},
        pages = {L25-L29},
          doi = {10.1086/184502},
       adsurl = {https://ui.adsabs.harvard.edu/abs/1985ApJ...294L..25D}
}

@ARTICLE{Douglas2022,
       author = {{Douglas}, Andrew and {Padmanabh}, Prajwal V. and {Ransom}, Scott M. and {Ridolfi}, Alessandro and {Freire}, Paulo and {Krishnan}, Vivek Venkatraman and {Barr}, Ewan D. and {Pallanca}, Cristina and {Cadelano}, Mario and {Possenti}, Andrea and {Stairs}, Ingrid and {Hessels}, Jason W.~T. and {DeCesar}, Megan E. and {Lynch}, Ryan S. and {Bailes}, Matthew and {Burgay}, Marta and {Champion}, David J. and {Karuppusamy}, Ramesh and {Kramer}, Michael and {Stappers}, Benjamin and {Vleeschower}, Laila},
        title = "{Two New Black Widow Millisecond Pulsars in M28}",
      journal = {\apj},
         year = 2022,
        month = mar,
       volume = {927},
       number = {1},
          eid = {126},
        pages = {126},
          doi = {10.3847/1538-4357/ac4744},
archivePrefix = {arXiv},
       eprint = {2201.11238},
 primaryClass = {astro-ph.HE},
       adsurl = {https://ui.adsabs.harvard.edu/abs/2022ApJ...927..126D}
}

@ARTICLE{Ridolfi2022,
       author = {{Ridolfi}, A. and {Freire}, P.~C.~C. and {Gautam}, T. and {Ransom}, S.~M. and {Barr}, E.~D. and {Buchner}, S. and {Burgay}, M. and {Abbate}, F. and {Venkatraman Krishnan}, V. and {Vleeschower}, L. and {Possenti}, A. and {Stappers}, B.~W. and {Kramer}, M. and {Chen}, W. and {Padmanabh}, P.~V. and {Champion}, D.~J. and {Bailes}, M. and {Levin}, L. and {Keane}, E.~F. and {Breton}, R.~P. and {Bezuidenhout}, M. and {Grie{\ss}meier}, J. -M. and {K{\"u}nkel}, L. and {Men}, Y. and {Camilo}, F. and {Geyer}, M. and {Hugo}, B.~V. and {Jameson}, A. and {Parthasarathy}, A. and {Serylak}, M.},
        title = "{TRAPUM discovery of 13 new pulsars in NGC 1851 using MeerKAT}",
      journal = {\aap},
         year = 2022,
        month = aug,
       volume = {664},
          eid = {A27},
        pages = {A27},
          doi = {10.1051/0004-6361/202143006},
archivePrefix = {arXiv},
       eprint = {2203.12302},
 primaryClass = {astro-ph.HE},
       adsurl = {https://ui.adsabs.harvard.edu/abs/2022A&A...664A..27R}
}

@ARTICLE{Vleeschower2022,
       author = {{Vleeschower}, L. and {Stappers}, B.~W. and {Bailes}, M. and {Barr}, E.~D. and {Kramer}, M. and {Ransom}, S. and {Ridolfi}, A. and {Venkatraman Krishnan}, V. and {Possenti}, A. and {Keith}, M.~J. and {Burgay}, M. and {Freire}, P.~C.~C. and {Spiewak}, R. and {Champion}, D.~J. and {Bezuidenhout}, M.~C. and {Ni{\c{t}}u}, I.~C. and {Chen}, W. and {Parthasarathy}, A. and {DeCesar}, M.~E. and {Buchner}, S. and {Stairs}, I.~H. and {Hessels}, J.~W.~T.},
        title = "{Discoveries and timing of pulsars in NGC 6440}",
      journal = {\mnras},
         year = 2022,
        month = jun,
       volume = {513},
       number = {1},
        pages = {1386-1399},
          doi = {10.1093/mnras/stac921},
archivePrefix = {arXiv},
       eprint = {2204.00086},
 primaryClass = {astro-ph.HE},
       adsurl = {https://ui.adsabs.harvard.edu/abs/2022MNRAS.513.1386V}
}

@ARTICLE{Abbate2022,
       author = {{Abbate}, F. and {Ridolfi}, A. and {Barr}, E.~D. and {Buchner}, S. and {Burgay}, M. and {Champion}, D.~J. and {Chen}, W. and {Freire}, P.~C.~C. and {Gautam}, T. and {Grie{\ss}meier}, J.~M. and {K{\"u}nkel}, L. and {Kramer}, M. and {Padmanabh}, P.~V. and {Possenti}, A. and {Ransom}, S. and {Serylak}, M. and {Stappers}, B.~W. and {Venkatraman Krishnan}, V. and {Behrend}, J. and {Breton}, R.~P. and {Levin}, L. and {Men}, Y.},
        title = "{Four pulsar discoveries in NGC 6624 by TRAPUM using MeerKAT}",
      journal = {\mnras},
         year = 2022,
        month = jun,
       volume = {513},
       number = {2},
        pages = {2292-2301},
          doi = {10.1093/mnras/stac1041},
archivePrefix = {arXiv},
       eprint = {2204.05334},
 primaryClass = {astro-ph.HE},
       adsurl = {https://ui.adsabs.harvard.edu/abs/2022MNRAS.513.2292A}
}

@ARTICLE{Freire2004,
       author = {{Freire}, Paulo C. and {Gupta}, Yashwant and {Ransom}, Scott M. and {Ishwara-Chandra}, C.~H.},
        title = "{Giant Metrewave Radio Telescope Discovery of a Millisecond Pulsar in a Very Eccentric Binary System}",
      journal = {\apjl},
         year = 2004,
        month = may,
       volume = {606},
       number = {1},
        pages = {L53-L56},
          doi = {10.1086/421085},
archivePrefix = {arXiv},
       eprint = {astro-ph/0403453},
 primaryClass = {astro-ph},
       adsurl = {https://ui.adsabs.harvard.edu/abs/2004ApJ...606L..53F}
}

@ARTICLE{Clark1975,
       author = {{Clark}, G.~W.},
        title = "{X-ray binaries in globular clusters.}",
      journal = {\apjl},
         year = 1975,
        month = aug,
       volume = {199},
        pages = {L143-L145},
          doi = {10.1086/181869},
       adsurl = {https://ui.adsabs.harvard.edu/abs/1975ApJ...199L.143C}
}

@INPROCEEDINGS{Verbunt1987,
       author = {{Verbunt}, F. and {Hut}, P.},
        title = "{The Globular Cluster Population of X-Ray Binaries}",
    booktitle = {The Origin and Evolution of Neutron Stars},
         year = 1987,
       editor = {{Helfand}, D.~J. and {Huang}, J. -H.},
       volume = {125},
        month = jan,
        pages = {187},
       adsurl = {https://ui.adsabs.harvard.edu/abs/1987IAUS..125..187V}
}

@ARTICLE{Ridolfi2016,
       author = {{Ridolfi}, A. and {Freire}, P.~C.~C. and {Torne}, P. and {Heinke}, C.~O. and {van den Berg}, M. and {Jordan}, C. and {Kramer}, M. and {Bassa}, C.~G. and {Sarkissian}, J. and {D'Amico}, N. and {Lorimer}, D. and {Camilo}, F. and {Manchester}, R.~N. and {Lyne}, A.},
        title = "{Long-term observations of the pulsars in 47 Tucanae - I. A study of four elusive binary systems}",
      journal = {\mnras},
         year = 2016,
        month = nov,
       volume = {462},
       number = {3},
        pages = {2918-2933},
          doi = {10.1093/mnras/stw1850},
archivePrefix = {arXiv},
       eprint = {1607.07248},
 primaryClass = {astro-ph.HE},
       adsurl = {https://ui.adsabs.harvard.edu/abs/2016MNRAS.462.2918R}
}

@ARTICLE{Abbate2018,
       author = {{Abbate}, F. and {Possenti}, A. and {Ridolfi}, A. and {Freire}, P.~C.~C. and {Camilo}, F. and {Manchester}, R.~N. and {D'Amico}, N.},
        title = "{Internal gas models and central black hole in 47 Tucanae using millisecond pulsars}",
      journal = {\mnras},
         year = 2018,
        month = nov,
       volume = {481},
       number = {1},
        pages = {627-638},
          doi = {10.1093/mnras/sty2298},
archivePrefix = {arXiv},
       eprint = {1808.06621},
 primaryClass = {astro-ph.HE},
       adsurl = {https://ui.adsabs.harvard.edu/abs/2018MNRAS.481..627A}
}

@ARTICLE{Ransom2005,
       author = {{Ransom}, Scott M. and {Hessels}, Jason W.~T. and {Stairs}, Ingrid H. and {Freire}, Paulo C.~C. and {Camilo}, Fernando and {Kaspi}, Victoria M. and {Kaplan}, David L.},
        title = "{Twenty-One Millisecond Pulsars in Terzan 5 Using the Green Bank Telescope}",
      journal = {Science},
         year = 2005,
        month = feb,
       volume = {307},
       number = {5711},
        pages = {892-896},
          doi = {10.1126/science.1108632},
archivePrefix = {arXiv},
       eprint = {astro-ph/0501230},
 primaryClass = {astro-ph},
       adsurl = {https://ui.adsabs.harvard.edu/abs/2005Sci...307..892R}
}

@ARTICLE{Freire2001,
       author = {{Freire}, P.~C. and {Camilo}, F. and {Lorimer}, D.~R. and {Lyne}, A.~G. and {Manchester}, R.~N. and {D'Amico}, N.},
        title = "{Timing the millisecond pulsars in 47 Tucanae}",
      journal = {\mnras},
         year = 2001,
        month = sep,
       volume = {326},
       number = {3},
        pages = {901-915},
          doi = {10.1046/j.1365-8711.2001.04493.x},
archivePrefix = {arXiv},
       eprint = {astro-ph/0103372},
 primaryClass = {astro-ph},
       adsurl = {https://ui.adsabs.harvard.edu/abs/2001MNRAS.326..901F}
}

@ARTICLE{Manchester1990,
       author = {{Manchester}, R.~N. and {Lyne}, A.~G. and {D'Amico}, N. and {Johnston}, S. and {Lim}, J. and {Kniffen}, D.~A.},
        title = "{A 5.75-millisecond pulsar in the globular cluster 47 Tucanae}",
      journal = {\nat},
         year = 1990,
        month = jun,
       volume = {345},
       number = {6276},
        pages = {598-600},
          doi = {10.1038/345598a0},
       adsurl = {https://ui.adsabs.harvard.edu/abs/1990Natur.345..598M}
}

@ARTICLE{Manchester1991,
       author = {{Manchester}, R.~N. and {Lyne}, A.~G. and {Robinson}, C. and {D'Amico}, N. and {Bailes}, M. and {Lim}, J.},
        title = "{Discovery of ten millisecond pulsars in the globular cluster 47 Tucanae}",
      journal = {\nat},
         year = 1991,
        month = jul,
       volume = {352},
       number = {6332},
        pages = {219-221},
          doi = {10.1038/352219a0},
       adsurl = {https://ui.adsabs.harvard.edu/abs/1991Natur.352..219M}
}

@ARTICLE{Robinson1995,
       author = {{Robinson}, Clive and {Lyne}, A.~G. and {Manchester}, R.~N. and {Bailes}, M. and {D'Amico}, N. and {Johnston}, S.},
        title = "{Millisecond pulsars in the globular cluster 47 Tucanae}",
      journal = {\mnras},
         year = 1995,
        month = may,
       volume = {274},
       number = {2},
        pages = {547-554},
          doi = {10.1093/mnras/274.2.547},
       adsurl = {https://ui.adsabs.harvard.edu/abs/1995MNRAS.274..547R}
}

@ARTICLE{Camilo2000,
       author = {{Camilo}, F. and {Lorimer}, D.~R. and {Freire}, P. and {Lyne}, A.~G. and {Manchester}, R.~N.},
        title = "{Observations of 20 Millisecond Pulsars in 47 Tucanae at 20 Centimeters}",
      journal = {\apj},
         year = 2000,
        month = jun,
       volume = {535},
       number = {2},
        pages = {975-990},
          doi = {10.1086/308859},
archivePrefix = {arXiv},
       eprint = {astro-ph/9911234},
 primaryClass = {astro-ph},
       adsurl = {https://ui.adsabs.harvard.edu/abs/2000ApJ...535..975C}
}

@ARTICLE{Freire2017,
       author = {{Freire}, P.~C.~C. and {Ridolfi}, A. and {Kramer}, M. and {Jordan}, C. and {Manchester}, R.~N. and {Torne}, P. and {Sarkissian}, J. and {Heinke}, C.~O. and {D'Amico}, N. and {Camilo}, F. and {Lorimer}, D.~R. and {Lyne}, A.~G.},
        title = "{Long-term observations of the pulsars in 47 Tucanae - II. Proper motions, accelerations and jerks}",
      journal = {\mnras},
         year = 2017,
        month = oct,
       volume = {471},
       number = {1},
        pages = {857-876},
          doi = {10.1093/mnras/stx1533},
archivePrefix = {arXiv},
       eprint = {1706.04908},
 primaryClass = {astro-ph.HE},
       adsurl = {https://ui.adsabs.harvard.edu/abs/2017MNRAS.471..857F}
}

@ARTICLE{Pan2016,
       author = {{Pan}, Z. and {Hobbs}, G. and {Li}, D. and {Ridolfi}, A. and {Wang}, P. and {Freire}, P.},
        title = "{Discovery of two new pulsars in 47 Tucanae (NGC 104)}",
      journal = {\mnras},
         year = 2016,
        month = jun,
       volume = {459},
       number = {1},
        pages = {L26-L30},
          doi = {10.1093/mnrasl/slw037},
archivePrefix = {arXiv},
       eprint = {1603.01348},
 primaryClass = {astro-ph.HE},
       adsurl = {https://ui.adsabs.harvard.edu/abs/2016MNRAS.459L..26P}
}

@ARTICLE{Chen2023,
       author = {{Chen}, W. and {Freire}, P.~C.~C. and {Ridolfi}, A. and {Barr}, E.~D. and {Stappers}, B. and {Kramer}, M. and {Possenti}, A. and {Ransom}, S.~M. and {Levin}, L. and {Breton}, R.~P. and {Burgay}, M. and {Camilo}, F. and {Buchner}, S. and {Champion}, D.~J. and {Abbate}, F. and {Venkatraman Krishnan}, V. and {Padmanabh}, P.~V. and {Gautam}, T. and {Vleeschower}, L. and {Geyer}, M. and {Grie{\ss}meier}, J. -M. and {Men}, Y.~P. and {Balakrishnan}, V. and {Bezuidenhout}, M.~C.},
        title = "{MeerKAT discovery of 13 new pulsars in Omega Centauri}",
      journal = {\mnras},
         year = 2023,
        month = apr,
       volume = {520},
       number = {3},
        pages = {3847-3856},
          doi = {10.1093/mnras/stad029},
archivePrefix = {arXiv},
       eprint = {2301.03864},
 primaryClass = {astro-ph.HE},
       adsurl = {https://ui.adsabs.harvard.edu/abs/2023MNRAS.520.3847C}
}

@ARTICLE{Abbate2023a,
       author = {{Abbate}, F. and {Possenti}, A. and {Ridolfi}, A. and {Venkatraman Krishnan}, V. and {Buchner}, S. and {Barr}, E.~D. and {Bailes}, M. and {Kramer}, M. and {Cameron}, A. and {Parthasarathy}, A. and {van Straten}, W. and {Chen}, W. and {Camilo}, F. and {Padmanabh}, P.~V. and {Mao}, S.~A. and {Freire}, P.~C.~C. and {Ransom}, S.~M. and {Vleeschower}, L. and {Geyer}, M. and {Zhang}, L.},
        title = "{A MeerKAT look at the polarization of 47 Tucanae pulsars: magnetic field implications}",
      journal = {\mnras},
         year = 2023,
        month = jan,
       volume = {518},
       number = {2},
        pages = {1642-1655},
          doi = {10.1093/mnras/stac3248},
archivePrefix = {arXiv},
       eprint = {2211.03815},
 primaryClass = {astro-ph.HE},
       adsurl = {https://ui.adsabs.harvard.edu/abs/2023MNRAS.518.1642A}
}

@ARTICLE{Abbate2023b,
       author = {{Abbate}, F. and {Ridolfi}, A. and {Freire}, P.~C.~C. and {Padmanabh}, P.~V. and {Balakrishnan}, V. and {Buchner}, S. and {Zhang}, L. and {Kramer}, M. and {Stappers}, B.~W. and {Barr}, E.~D. and {Chen}, W. and {Champion}, D. and {Ransom}, S. and {Possenti}, A.},
        title = "{A MeerKAT view of the pulsars in the globular cluster NGC 6522}",
      journal = {\aap},
         year = 2023,
        month = dec,
       volume = {680},
          eid = {A47},
        pages = {A47},
          doi = {10.1051/0004-6361/202347725},
archivePrefix = {arXiv},
       eprint = {2310.03800},
 primaryClass = {astro-ph.HE},
       adsurl = {https://ui.adsabs.harvard.edu/abs/2023A&A...680A..47A}
}

@ARTICLE{Men2023,
       author = {{Men}, Yunpeng and {Barr}, Ewan and {Clark}, Colin J. and {Carli}, Emma and {Desvignes}, Gregory},
        title = "{PulsarX: A new pulsar searching package. I. A high performance folding program for pulsar surveys}",
      journal = {\aap},
         year = 2023,
        month = nov,
       volume = {679},
          eid = {A20},
        pages = {A20},
          doi = {10.1051/0004-6361/202347356},
archivePrefix = {arXiv},
       eprint = {2309.02544},
 primaryClass = {astro-ph.IM},
       adsurl = {https://ui.adsabs.harvard.edu/abs/2023A&A...679A..20M}
}

@software{Ransom2011,
       author = {{Ransom}, Scott},
        title = "{PRESTO: PulsaR Exploration and Search TOolkit}",
 howpublished = {Astrophysics Source Code Library, record ascl:1107.017},
         year = 2011,
        month = jul,
          eid = {ascl:1107.017},
       adsurl = {https://ui.adsabs.harvard.edu/abs/2011ascl.soft07017R}
}

@ARTICLE{Hobbs2006,
       author = {{Hobbs}, G.~B. and {Edwards}, R.~T. and {Manchester}, R.~N.},
        title = "{TEMPO2, a new pulsar-timing package - I. An overview}",
      journal = {\mnras},
         year = 2006,
        month = jun,
       volume = {369},
       number = {2},
        pages = {655-672},
          doi = {10.1111/j.1365-2966.2006.10302.x},
archivePrefix = {arXiv},
       eprint = {astro-ph/0603381},
 primaryClass = {astro-ph},
       adsurl = {https://ui.adsabs.harvard.edu/abs/2006MNRAS.369..655H}
}

@ARTICLE{Bezuidenhout2023,
       author = {{Bezuidenhout}, M.~C. and {Clark}, C.~J. and {Breton}, R.~P. and {Stappers}, B.~W. and {Barr}, E.~D. and {Caleb}, M. and {Chen}, W. and {Jankowski}, F. and {Kramer}, M. and {Rajwade}, K. and {Surnis}, M.},
        title = "{Tied-array beam localization of radio transients and pulsars}",
      journal = {RAS Techniques and Instruments},
         year = 2023,
        month = jan,
       volume = {2},
       number = {1},
        pages = {114-128},
          doi = {10.1093/rasti/rzad007},
archivePrefix = {arXiv},
       eprint = {2302.09812},
 primaryClass = {astro-ph.HE},
       adsurl = {https://ui.adsabs.harvard.edu/abs/2023RASTI...2..114B}
}

@ARTICLE{Heywood2023,
       author = {{Heywood}, Ian},
        title = "{A new pulsar candidate in 47 Tucanae discovered with MeerKAT imaging}",
      journal = {\mnras},
         year = 2023,
        month = oct,
       volume = {525},
       number = {1},
        pages = {L76-L81},
          doi = {10.1093/mnrasl/slad094},
archivePrefix = {arXiv},
       eprint = {2307.02077},
 primaryClass = {astro-ph.HE},
       adsurl = {https://ui.adsabs.harvard.edu/abs/2023MNRAS.525L..76H}
}

@ARTICLE{Barr2024,
       author = {{Barr}, Ewan D. and {Dutta}, Arunima and {Freire}, Paulo C.~C. and {Cadelano}, Mario and {Gautam}, Tasha and {Kramer}, Michael and {Pallanca}, Cristina and {Ransom}, Scott M. and {Ridolfi}, Alessandro and {Stappers}, Benjamin W. and {Tauris}, Thomas M. and {Venkatraman Krishnan}, Vivek and {Wex}, Norbert and {Bailes}, Matthew and {Behrend}, Jan and {Buchner}, Sarah and {Burgay}, Marta and {Chen}, Weiwei and {Champion}, David J. and {Chen}, C. -H. Rosie and {Corongiu}, Alessandro and {Geyer}, Marisa and {Men}, Y.~P. and {Padmanabh}, Prajwal Voraganti and {Possenti}, Andrea},
        title = "{A pulsar in a binary with a compact object in the mass gap between neutron stars and black holes}",
      journal = {Science},
         year = 2024,
        month = jan,
       volume = {383},
       number = {6680},
        pages = {275-279},
          doi = {10.1126/science.adg3005},
archivePrefix = {arXiv},
       eprint = {2401.09872},
 primaryClass = {astro-ph.HE},
       adsurl = {https://ui.adsabs.harvard.edu/abs/2024Sci...383..275B}
}

@ARTICLE{Vleeschower2024,
       author = {{Vleeschower}, L. and {Corongiu}, A. and {Stappers}, B.~W. and {Freire}, P.~C.~C. and {Ridolfi}, A. and {Abbate}, F. and {Ransom}, S.~M. and {Possenti}, A. and {Padmanabh}, P.~V. and {Balakrishnan}, V. and {Kramer}, M. and {Venkatraman Krishnan}, V. and {Zhang}, L. and {Bailes}, M. and {Barr}, E.~D. and {Buchner}, S. and {Chen}, W.},
        title = "{Discoveries and timing of pulsars in M62}",
      journal = {\mnras},
         year = 2024,
        month = may,
       volume = {530},
       number = {2},
        pages = {1436-1456},
          doi = {10.1093/mnras/stae816},
archivePrefix = {arXiv},
       eprint = {2403.12137},
 primaryClass = {astro-ph.HE},
       adsurl = {https://ui.adsabs.harvard.edu/abs/2024MNRAS.530.1436V}
}

@ARTICLE{Padmanabh2024,
       author = {{Padmanabh}, P.~V. and {Ransom}, S.~M. and {Freire}, P.~C.~C. and {Ridolfi}, A. and {Taylor}, J.~D. and {Choza}, C. and {Clark}, C.~J. and {Abbate}, F. and {Bailes}, M. and {Barr}, E.~D. and {Buchner}, S. and {Burgay}, M. and {DeCesar}, M.~E. and {Chen}, W. and {Corongiu}, A. and {Champion}, D.~J. and {Dutta}, A. and {Geyer}, M. and {Hessels}, J.~W.~T. and {Kramer}, M. and {Possenti}, A. and {Stairs}, I.~H. and {Stappers}, B.~W. and {Venkatraman Krishnan}, V. and {Vleeschower}, L. and {Zhang}, L.},
        title = "{Discovery and timing of ten new millisecond pulsars in the globular cluster Terzan 5}",
      journal = {\aap},
         year = 2024,
        month = jun,
       volume = {686},
          eid = {A166},
        pages = {A166},
          doi = {10.1051/0004-6361/202449303},
archivePrefix = {arXiv},
       eprint = {2403.17799},
 primaryClass = {astro-ph.HE},
       adsurl = {https://ui.adsabs.harvard.edu/abs/2024A&A...686A.166P}
}

@ARTICLE{Kramer2021,
       author = {{Kramer}, M. and {Stairs}, I.~H. and {Manchester}, R.~N. and {Wex}, N. and {Deller}, A.~T. and {Coles}, W.~A. and {Ali}, M. and {Burgay}, M. and {Camilo}, F. and {Cognard}, I. and {Damour}, T. and {Desvignes}, G. and {Ferdman}, R.~D. and {Freire}, P.~C.~C. and {Grondin}, S. and {Guillemot}, L. and {Hobbs}, G.~B. and {Janssen}, G. and {Karuppusamy}, R. and {Lorimer}, D.~R. and {Lyne}, A.~G. and {McKee}, J.~W. and {McLaughlin}, M. and {M{\"u}nch}, L.~E. and {Perera}, B.~B.~P. and {Pol}, N. and {Possenti}, A. and {Sarkissian}, J. and {Stappers}, B.~W. and {Theureau}, G.},
        title = "{Strong-Field Gravity Tests with the Double Pulsar}",
      journal = {Physical Review X},
         year = 2021,
        month = oct,
       volume = {11},
       number = {4},
          eid = {041050},
        pages = {041050},
          doi = {10.1103/PhysRevX.11.041050},
archivePrefix = {arXiv},
       eprint = {2112.06795},
 primaryClass = {astro-ph.HE},
       adsurl = {https://ui.adsabs.harvard.edu/abs/2021PhRvX..11d1050K}
}

@ARTICLE{FreireWex2024,
       author = {{Freire}, Paulo C.~C. and {Wex}, Norbert},
        title = "{Gravity experiments with radio pulsars}",
      journal = {Living Reviews in Relativity},
         year = 2024,
        month = dec,
       volume = {27},
       number = {1},
          eid = {5},
        pages = {5},
          doi = {10.1007/s41114-024-00051-y},
archivePrefix = {arXiv},
       eprint = {2407.16540},
 primaryClass = {gr-qc},
       adsurl = {https://ui.adsabs.harvard.edu/abs/2024LRR....27....5F}
}

@ARTICLE{Voisin2020,
       author = {{Voisin}, G. and {Cognard}, I. and {Freire}, P.~C.~C. and {Wex}, N. and {Guillemot}, L. and {Desvignes}, G. and {Kramer}, M. and {Theureau}, G.},
        title = "{An improved test of the strong equivalence principle with the pulsar in a triple star system}",
      journal = {\aap},
         year = 2020,
        month = jun,
       volume = {638},
          eid = {A24},
        pages = {A24},
          doi = {10.1051/0004-6361/202038104},
archivePrefix = {arXiv},
       eprint = {2005.01388},
 primaryClass = {gr-qc},
       adsurl = {https://ui.adsabs.harvard.edu/abs/2020A&A...638A..24V}
}

@ARTICLE{OzelFreire2016,
       author = {{{\"O}zel}, Feryal and {Freire}, Paulo},
        title = "{Masses, Radii, and the Equation of State of Neutron Stars}",
      journal = {\araa},
         year = 2016,
        month = sep,
       volume = {54},
        pages = {401-440},
          doi = {10.1146/annurev-astro-081915-023322},
archivePrefix = {arXiv},
       eprint = {1603.02698},
 primaryClass = {astro-ph.HE},
       adsurl = {https://ui.adsabs.harvard.edu/abs/2016ARA&A..54..401O}
}

@ARTICLE{Fonseca2021,
       author = {{Fonseca}, E. and {Cromartie}, H.~T. and {Pennucci}, T.~T. and {Ray}, P.~S. and {Kirichenko}, A. Yu. and {Ransom}, S.~M. and {Demorest}, P.~B. and {Stairs}, I.~H. and {Arzoumanian}, Z. and {Guillemot}, L. and {Parthasarathy}, A. and {Kerr}, M. and {Cognard}, I. and {Baker}, P.~T. and {Blumer}, H. and {Brook}, P.~R. and {DeCesar}, M. and {Dolch}, T. and {Dong}, F.~A. and {Ferrara}, E.~C. and {Fiore}, W. and {Garver-Daniels}, N. and {Good}, D.~C. and {Jennings}, R. and {Jones}, M.~L. and {Kaspi}, V.~M. and {Lam}, M.~T. and {Lorimer}, D.~R. and {Luo}, J. and {McEwen}, A. and {McKee}, J.~W. and {McLaughlin}, M.~A. and {McMann}, N. and {Meyers}, B.~W. and {Naidu}, A. and {Ng}, C. and {Nice}, D.~J. and {Pol}, N. and {Radovan}, H.~A. and {Shapiro-Albert}, B. and {Tan}, C.~M. and {Tendulkar}, S.~P. and {Swiggum}, J.~K. and {Wahl}, H.~M. and {Zhu}, W.~W.},
        title = "{Refined Mass and Geometric Measurements of the High-mass PSR J0740+6620}",
      journal = {\apjl},
         year = 2021,
        month = jul,
       volume = {915},
       number = {1},
          eid = {L12},
        pages = {L12},
          doi = {10.3847/2041-8213/ac03b8},
archivePrefix = {arXiv},
       eprint = {2104.00880},
 primaryClass = {astro-ph.HE},
       adsurl = {https://ui.adsabs.harvard.edu/abs/2021ApJ...915L..12F}
}

@BOOK{Tauris2023,
       author = {{Tauris}, Thomas M. and {van den Heuvel}, Edward P.~J.},
        title = "{Physics of Binary Star Evolution. From Stars to X-ray Binaries and Gravitational Wave Sources}",
         year = 2023,
          doi = {10.48550/arXiv.2305.09388},
       adsurl = {https://ui.adsabs.harvard.edu/abs/2023pbse.book.....T}
}

@ARTICLE{Kramer2021b,
       author = {{Kramer}, M. and {Stairs}, I.~H. and {Venkatraman Krishnan}, V. and {Freire}, P.~C.~C. and {Abbate}, F. and {Bailes}, M. and {Burgay}, M. and {Buchner}, S. and {Champion}, D.~J. and {Cognard}, I. and {Gautam}, T. and {Geyer}, M. and {Guillemot}, L. and {Hu}, H. and {Janssen}, G. and {Lower}, M.~E. and {Parthasarathy}, A. and {Possenti}, A. and {Ransom}, S. and {Reardon}, D.~J. and {Ridolfi}, A. and {Serylak}, M. and {Shannon}, R.~M. and {Spiewak}, R. and {Theureau}, G. and {van Straten}, W. and {Wex}, N. and {Oswald}, L.~S. and {Posselt}, B. and {Sobey}, C. and {Barr}, E.~D. and {Camilo}, F. and {Hugo}, B. and {Jameson}, A. and {Johnston}, S. and {Karastergiou}, A. and {Keith}, M. and {Os{\l}owski}, S.},
        title = "{The relativistic binary programme on MeerKAT: science objectives and first results}",
      journal = {\mnras},
         year = 2021,
        month = jun,
       volume = {504},
       number = {2},
        pages = {2094-2114},
          doi = {10.1093/mnras/stab375},
archivePrefix = {arXiv},
       eprint = {2102.05160},
 primaryClass = {astro-ph.HE},
       adsurl = {https://ui.adsabs.harvard.edu/abs/2021MNRAS.504.2094K}
}

@ARTICLE{Edmonds2002,
       author = {{Edmonds}, Peter D. and {Gilliland}, Ronald L. and {Camilo}, Fernando and {Heinke}, Craig O. and {Grindlay}, Jonathan E.},
        title = "{A Millisecond Pulsar Optical Counterpart with Large-Amplitude Variability in the Globular Cluster 47 Tucanae}",
      journal = {\apj},
         year = 2002,
        month = nov,
       volume = {579},
       number = {2},
        pages = {741-751},
          doi = {10.1086/342985},
archivePrefix = {arXiv},
       eprint = {astro-ph/0207426},
 primaryClass = {astro-ph},
       adsurl = {https://ui.adsabs.harvard.edu/abs/2002ApJ...579..741E}
}

@ARTICLE{Grindlay2001,
       author = {{Grindlay}, Jonathan E. and {Heinke}, Craig and {Edmonds}, Peter D. and {Murray}, Stephen S.},
        title = "{High-Resolution X-ray Imaging of a Globular Cluster Core: Compact Binaries in 47Tuc}",
      journal = {Science},
         year = 2001,
        month = jun,
       volume = {292},
       number = {5525},
        pages = {2290-2295},
          doi = {10.1126/science.1061135},
archivePrefix = {arXiv},
       eprint = {astro-ph/0105528},
 primaryClass = {astro-ph},
       adsurl = {https://ui.adsabs.harvard.edu/abs/2001Sci...292.2290G}
}

@ARTICLE{Bahramian2013,
       author = {{Bahramian}, Arash and {Heinke}, Craig O. and {Sivakoff}, Gregory R. and {Gladstone}, Jeanette C.},
        title = "{Stellar Encounter Rate in Galactic Globular Clusters}",
      journal = {\apj},
         year = 2013,
        month = apr,
       volume = {766},
       number = {2},
          eid = {136},
        pages = {136},
          doi = {10.1088/0004-637X/766/2/136},
archivePrefix = {arXiv},
       eprint = {1302.2549},
 primaryClass = {astro-ph.HE},
       adsurl = {https://ui.adsabs.harvard.edu/abs/2013ApJ...766..136B}
}

@ARTICLE{Freire2003,
       author = {{Freire}, P.~C. and {Camilo}, F. and {Kramer}, M. and {Lorimer}, D.~R. and {Lyne}, A.~G. and {Manchester}, R.~N. and {D'Amico}, N.},
        title = "{Further results from the timing of the millisecond pulsars in 47 Tucanae}",
      journal = {\mnras},
         year = 2003,
        month = apr,
       volume = {340},
       number = {4},
        pages = {1359-1374},
          doi = {10.1046/j.1365-8711.2003.06392.x},
       adsurl = {https://ui.adsabs.harvard.edu/abs/2003MNRAS.340.1359F}
}

@ARTICLE{Rivera-Sandoval2015,
       author = {{Rivera-Sandoval}, L.~E. and {van den Berg}, M. and {Heinke}, C.~O. and {Cohn}, H.~N. and {Lugger}, P.~M. and {Freire}, P. and {Anderson}, J. and {Serenelli}, A.~M. and {Althaus}, L.~G. and {Cool}, A.~M. and {Grindlay}, J.~E. and {Edmonds}, P.~D. and {Wijnands}, R. and {Ivanova}, N.},
        title = "{Discovery of near-ultraviolet counterparts to millisecond pulsars in the globular cluster 47 Tucanae}",
      journal = {\mnras},
         year = 2015,
        month = nov,
       volume = {453},
       number = {3},
        pages = {2707-2717},
          doi = {10.1093/mnras/stv1810},
archivePrefix = {arXiv},
       eprint = {1508.05291},
 primaryClass = {astro-ph.SR},
       adsurl = {https://ui.adsabs.harvard.edu/abs/2015MNRAS.453.2707R}
}

@ARTICLE{Cadelano2015,
       author = {{Cadelano}, M. and {Pallanca}, C. and {Ferraro}, F.~R. and {Salaris}, M. and {Dalessandro}, E. and {Lanzoni}, B. and {Freire}, P.~C.~C.},
        title = "{Optical Identification of He White Dwarfs Orbiting Four Millisecond Pulsars in the Globular Cluster 47 Tucanae}",
      journal = {\apj},
         year = 2015,
        month = oct,
       volume = {812},
       number = {1},
          eid = {63},
        pages = {63},
          doi = {10.1088/0004-637X/812/1/63},
archivePrefix = {arXiv},
       eprint = {1509.01397},
 primaryClass = {astro-ph.SR},
       adsurl = {https://ui.adsabs.harvard.edu/abs/2015ApJ...812...63C}
}

@ARTICLE{Edmonds2001,
       author = {{Edmonds}, Peter D. and {Gilliland}, Ronald L. and {Heinke}, Craig O. and {Grindlay}, Jonathan E. and {Camilo}, Fernando},
        title = "{Optical Detection of a Variable Millisecond Pulsar Companion in 47 Tucanae}",
      journal = {\apjl},
         year = 2001,
        month = aug,
       volume = {557},
       number = {1},
        pages = {L57-L60},
          doi = {10.1086/323122},
archivePrefix = {arXiv},
       eprint = {astro-ph/0107096},
 primaryClass = {astro-ph},
       adsurl = {https://ui.adsabs.harvard.edu/abs/2001ApJ...557L..57E}
}

@ARTICLE{Freire2001b,
       author = {{Freire}, P.~C. and {Kramer}, M. and {Lyne}, A.~G. and {Camilo}, F. and {Manchester}, R.~N. and {D'Amico}, N.},
        title = "{Detection of Ionized Gas in the Globular Cluster 47 Tucanae}",
      journal = {\apjl},
         year = 2001,
        month = aug,
       volume = {557},
       number = {2},
        pages = {L105-L108},
          doi = {10.1086/323248},
archivePrefix = {arXiv},
       eprint = {astro-ph/0107206},
 primaryClass = {astro-ph},
       adsurl = {https://ui.adsabs.harvard.edu/abs/2001ApJ...557L.105F}
}

@ARTICLE{Bogdanov2005,
       author = {{Bogdanov}, Slavko and {Grindlay}, Jonathan E. and {van den Berg}, Maureen},
        title = "{An X-Ray Variable Millisecond Pulsar in the Globular Cluster 47 Tucanae: Closing the Link to Low-Mass X-Ray Binaries}",
      journal = {\apj},
         year = 2005,
        month = sep,
       volume = {630},
       number = {2},
        pages = {1029-1036},
          doi = {10.1086/432249},
archivePrefix = {arXiv},
       eprint = {astro-ph/0506031},
 primaryClass = {astro-ph},
       adsurl = {https://ui.adsabs.harvard.edu/abs/2005ApJ...630.1029B}
}

@ARTICLE{Bogdanov2006,
       author = {{Bogdanov}, Slavko and {Grindlay}, Jonathan E. and {Heinke}, Craig O. and {Camilo}, Fernando and {Freire}, Paulo C.~C. and {Becker}, Werner},
        title = "{Chandra X-Ray Observations of 19 Millisecond Pulsars in the Globular Cluster 47 Tucanae}",
      journal = {\apj},
         year = 2006,
        month = aug,
       volume = {646},
       number = {2},
        pages = {1104-1115},
          doi = {10.1086/505133},
archivePrefix = {arXiv},
       eprint = {astro-ph/0604318},
 primaryClass = {astro-ph},
       adsurl = {https://ui.adsabs.harvard.edu/abs/2006ApJ...646.1104B}
}

@ARTICLE{Bhattacharya2017,
       author = {{Bhattacharya}, Souradeep and {Heinke}, Craig O. and {Chugunov}, Andrey I. and {Freire}, Paulo C.~C. and {Ridolfi}, Alessandro and {Bogdanov}, Slavko},
        title = "{Chandra studies of the globular cluster 47 Tucanae: A deeper X-ray source catalogue, five new X-ray counterparts to millisecond radio pulsars, and new constraints to r-mode instability window}",
      journal = {\mnras},
         year = 2017,
        month = dec,
       volume = {472},
       number = {3},
        pages = {3706-3721},
          doi = {10.1093/mnras/stx2241},
archivePrefix = {arXiv},
       eprint = {1709.01807},
 primaryClass = {astro-ph.HE},
       adsurl = {https://ui.adsabs.harvard.edu/abs/2017MNRAS.472.3706B}
}

@ARTICLE{Hebbar2021,
       author = {{Hebbar}, P.~R. and {Heinke}, C.~O. and {Kandel}, D. and {Romani}, R.~W. and {Freire}, P.~C.~C.},
        title = "{On the vanishing orbital X-ray variability of the eclipsing binary millisecond pulsar 47 Tuc W}",
      journal = {\mnras},
         year = 2021,
        month = jan,
       volume = {500},
       number = {1},
        pages = {1139-1150},
          doi = {10.1093/mnras/staa3072},
archivePrefix = {arXiv},
       eprint = {2009.13561},
 primaryClass = {astro-ph.HE},
       adsurl = {https://ui.adsabs.harvard.edu/abs/2021MNRAS.500.1139H}
}

@ARTICLE{Shao2019,
       author = {{Shao}, Zhengyi and {Li}, Lu},
        title = "{Gaia parallax of Milky Way globular clusters - A solution of mixture model}",
      journal = {\mnras},
         year = 2019,
        month = nov,
       volume = {489},
       number = {3},
        pages = {3093-3101},
          doi = {10.1093/mnras/stz2317},
archivePrefix = {arXiv},
       eprint = {1908.06031},
 primaryClass = {astro-ph.GA},
       adsurl = {https://ui.adsabs.harvard.edu/abs/2019MNRAS.489.3093S}
}

@article{Risbud2026,
	author = {{Risbud, D.} and {Ridolfi, A.} and {Freire, P. C. C.} and {Cadelano, M.} and {Chen, W.} and {Zhang, L.} and {Nag, R.} and {Camilo, F.} and {Padmanabh, P. V.} and {Corongiu, A.} and {Abbate, F.} and {Possenti, A.}},
	title = {PSR J0024-7204ai: A massive eccentric binary system in the globular cluster 47 Tucanae},
	DOI= "10.1051/0004-6361/202558541",
	url= "https://doi.org/10.1051/0004-6361/202558541",
	journal = {A\&A},
	year = 2026,
	volume = 708,
	pages = "A158",
}

@ARTICLE{Chabrier2003,
       author = {{Chabrier}, Gilles},
        title = "{Galactic Stellar and Substellar Initial Mass Function}",
      journal = {\pasp},
         year = 2003,
        month = jul,
       volume = {115},
       number = {809},
        pages = {763-795},
          doi = {10.1086/376392},
archivePrefix = {arXiv},
       eprint = {astro-ph/0304382},
 primaryClass = {astro-ph},
       adsurl = {https://ui.adsabs.harvard.edu/abs/2003PASP..115..763C}
}

@ARTICLE{Baumgardt2018,
       author = {{Baumgardt}, H. and {Hilker}, M.},
        title = "{A catalogue of masses, structural parameters, and velocity dispersion profiles of 112 Milky Way globular clusters}",
      journal = {\mnras},
         year = 2018,
        month = aug,
       volume = {478},
       number = {2},
        pages = {1520-1557},
          doi = {10.1093/mnras/sty1057},
archivePrefix = {arXiv},
       eprint = {1804.08359},
 primaryClass = {astro-ph.GA},
       adsurl = {https://ui.adsabs.harvard.edu/abs/2018MNRAS.478.1520B}
}
\clearpage

\begin{appendix}
\section{Parkes radio telescope archival data re-detection summary }

In this appendix we show the brightest re-detection of the newly discovered pulsars in the archival data of the Parkes radio telescope. See Sect. \ref{sec:parkes_redetections} for details.   

\begin{figure}[h]
    \centering
    \includegraphics[width=\columnwidth]{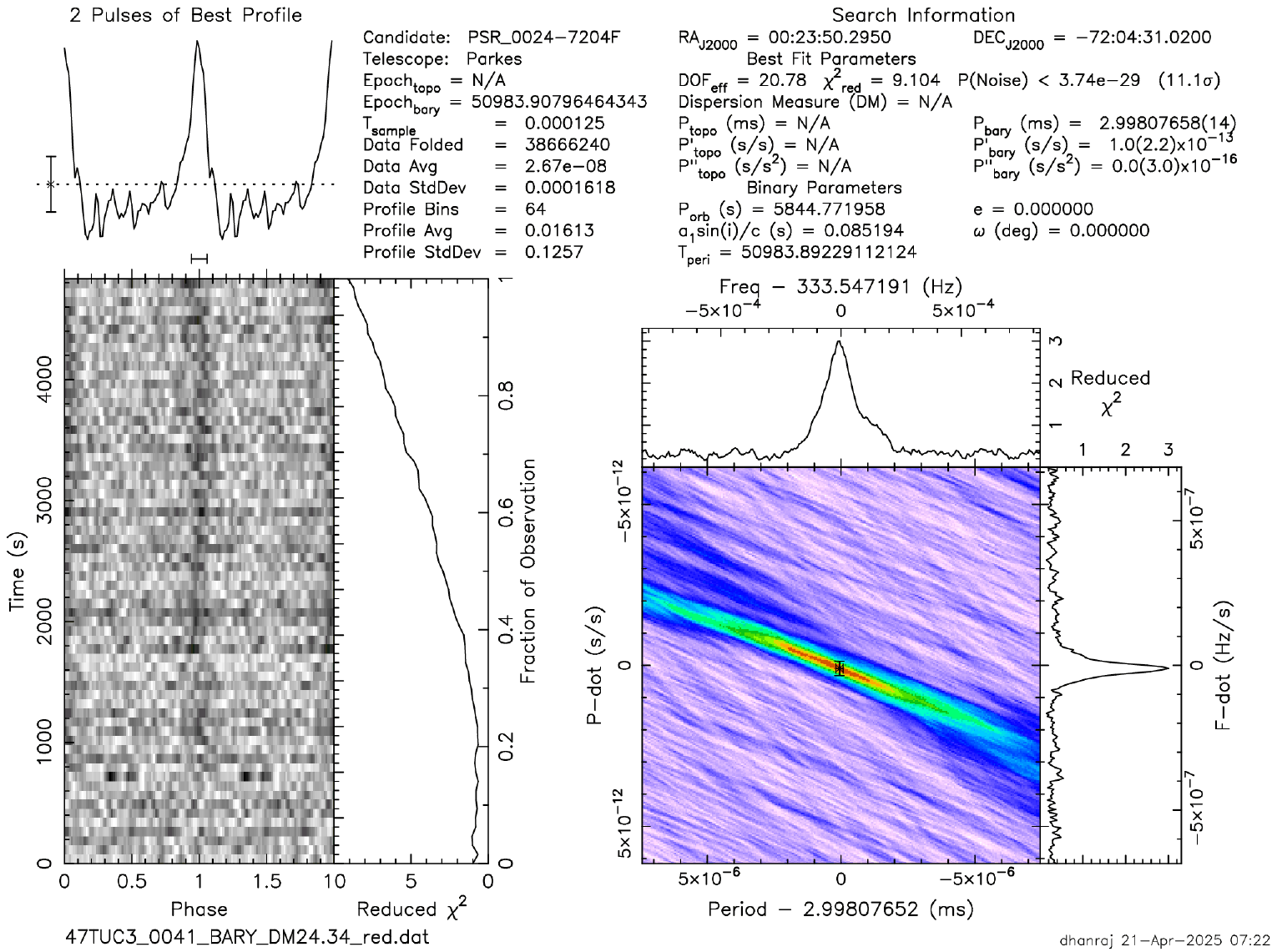}
    \caption{Re-detection of pulsar 47~Tuc~af. During the first $\sim$1000 sec of the observation, the pulsar signal is missing, as a combined effect of an eclipse (dominant) and unfavourable scintillation. }
    \label{fig:redetection_af}
\end{figure}

\begin{figure}[h]
    \centering
    \includegraphics[width=\columnwidth]{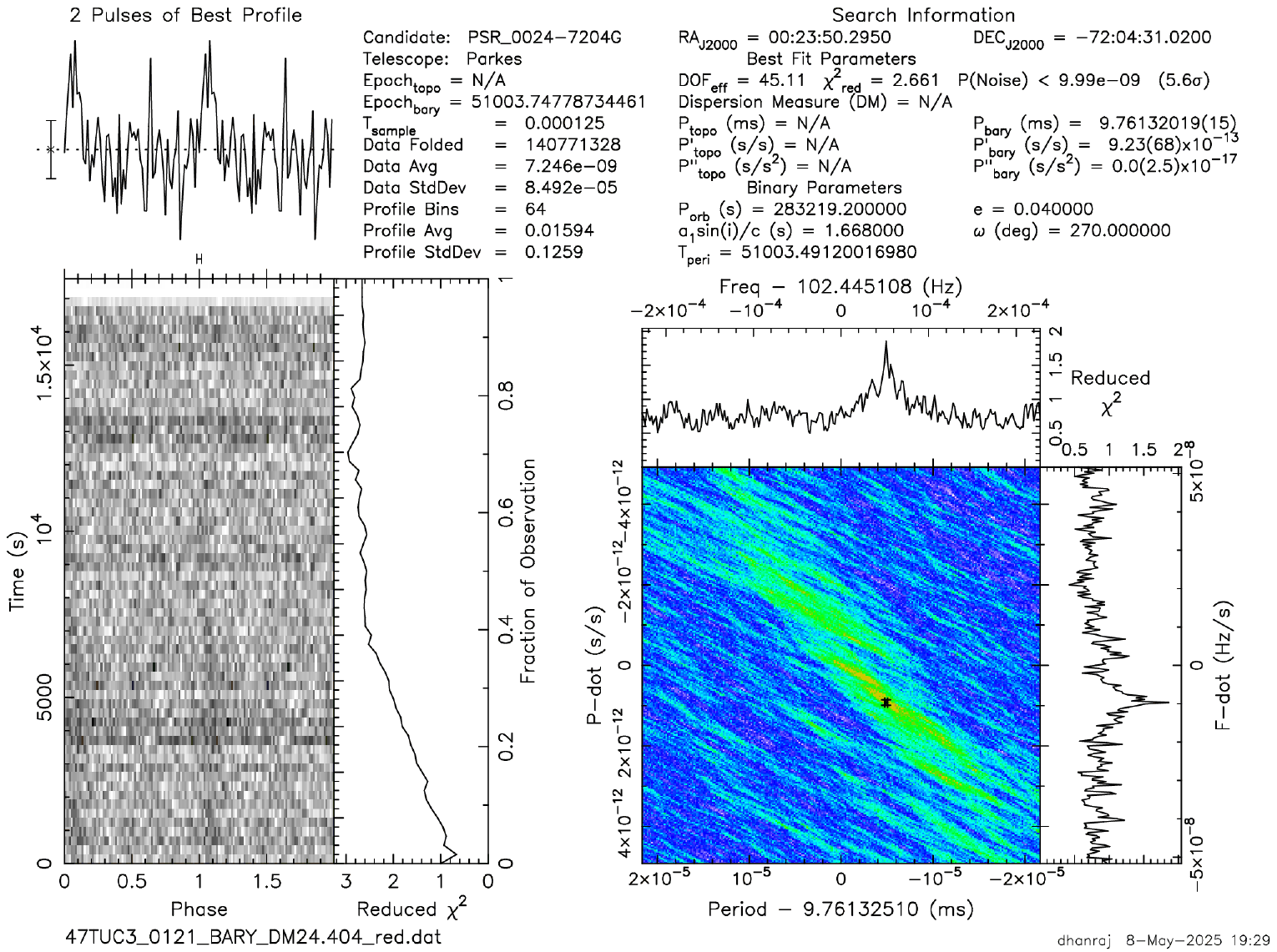}
    \caption{Re-detection of pulsar 47~Tuc~ag. The pulsar signal disappears after $\approx$ 10$^4$ sec as the scintillation varies to unfavourable scenario.  }
    \label{fig:redetection_ag}
\end{figure}

\begin{figure}[h]
    \centering
    \includegraphics[width=\columnwidth]{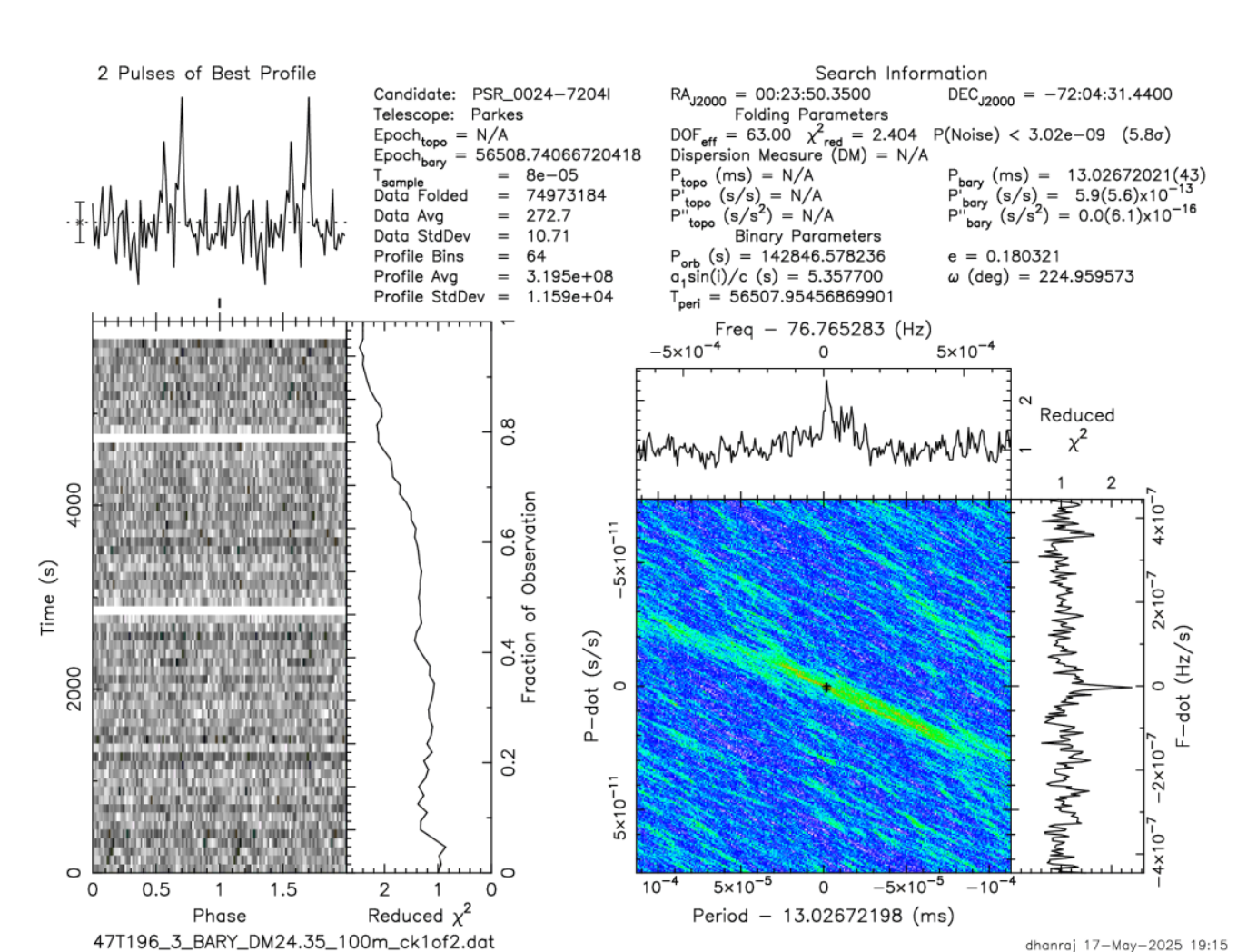}
    \caption{Re-detection of pulsar 47~Tuc~ai. Some sub-integrations with short-lived RFI instances have been zapped. }
    \label{fig:redetection_ai}
\end{figure}

\begin{figure}[h]
    \centering
    \includegraphics[width=\columnwidth]{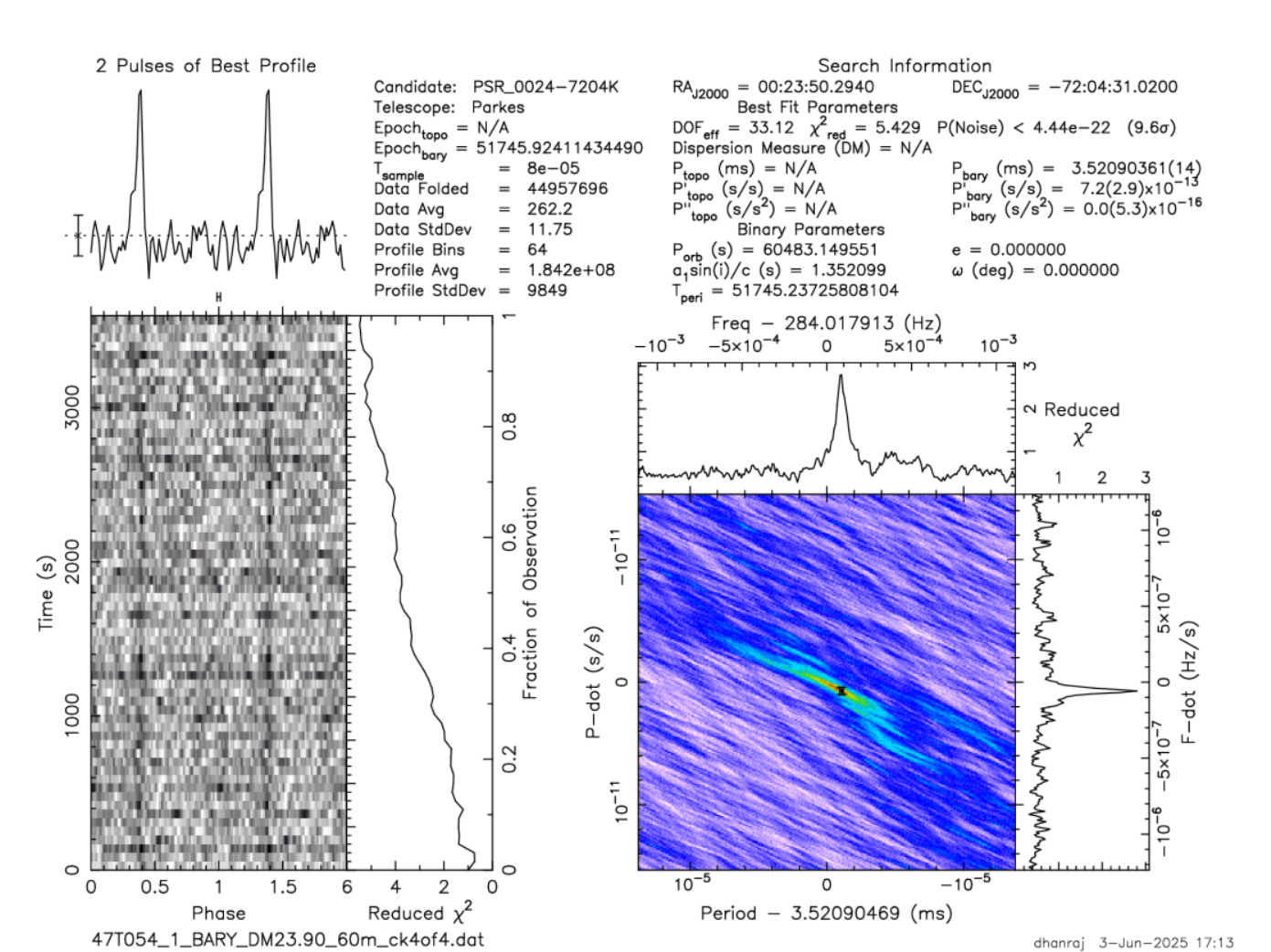}
    \caption{Re-detection of pulsar 47~Tuc~ak. A slight profile variation is seen compared to the UHF profile shown in Fig. \ref{fig:profiles_of_new_pulsars}. }
    \label{fig:redetection_ak}
\end{figure}

\begin{figure}[h]
    \centering
    \includegraphics[width=\columnwidth]{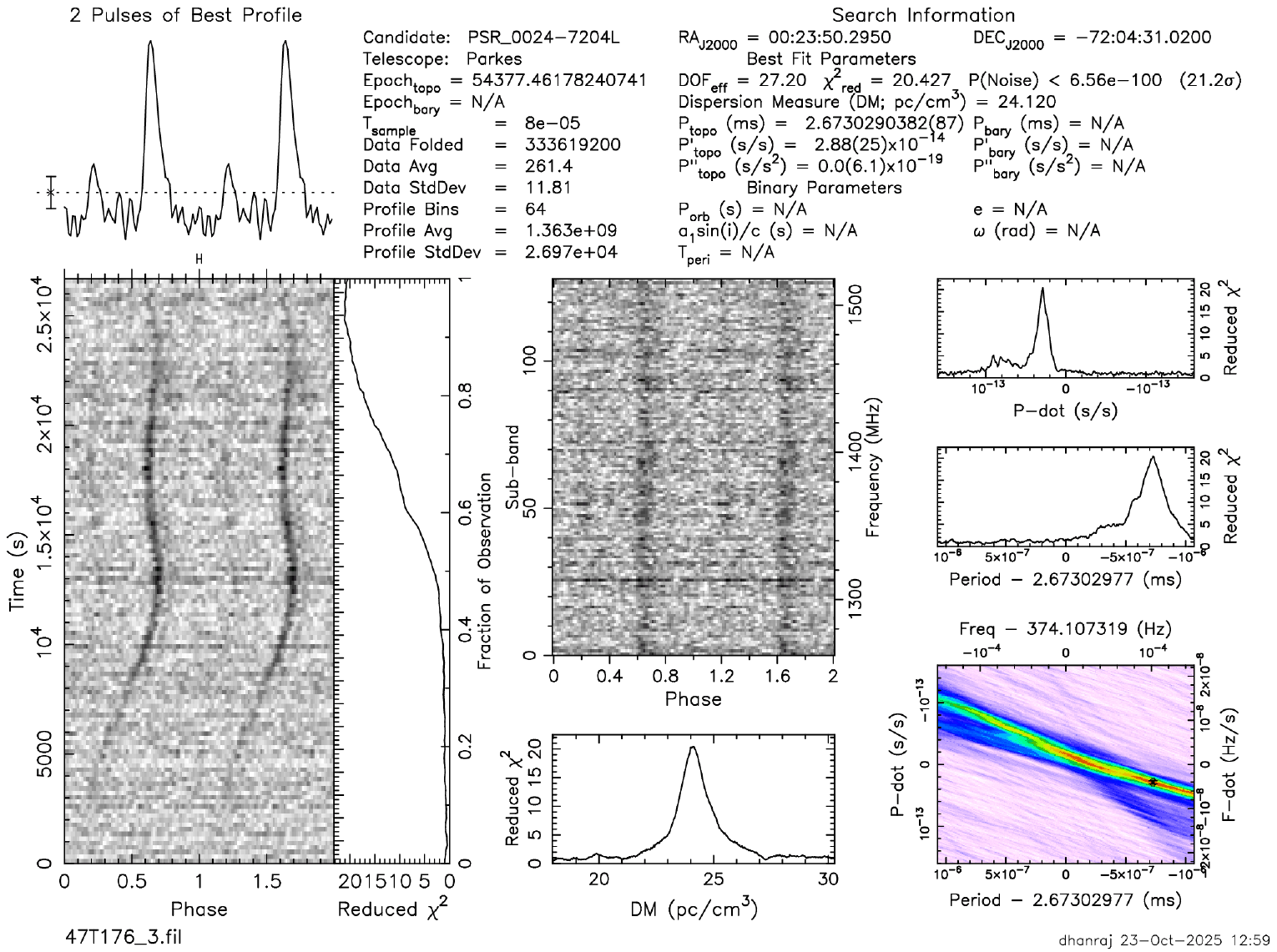}
    \caption{Re-detection of pulsar 47~Tuc~al. although the orbital parameters derived from MeerKAT detections were good enough to detect the pulsar after folding with \textsc{spider\_twister}, they are still slightly inaccurate, which possibly caused the pulsar signal to drift in phase. Alternatively, it could also be due to the orbital variability, typically seen in the BW systems. }
    \label{fig:redetection_al}
\end{figure}

 \end{appendix}

\end{document}